# STUDIES OF THE DIFFUSE INTERSTELLAR BANDS. III. HD 183143

L. M. Hobbs[1], D. G. York[2,3], J. A. Thorburn[1,10], T. P. Snow[4], M. Bishof[2], S. D. Friedman[5],.B. J. McCall[6], T. Oka[2,3], B. Rachford[7], P. Sonnentrucker[8], and D. E. Welty[9]

## ABSTRACT

Echelle spectra of HD 183143 [B7Iae, E(B-V) = 1.27] were obtained on three nights, at a resolving power R = 38,000 and with a S/N ratio ≈ 1000 at 6400 Å in the final, combined spectrum. A catalog is presented of 414 diffuse interstellar bands (DIBs) measured between 3900 and 8100 Å in this spectrum. The central wavelengths, the widths (FWHM), and the equivalent widths of nearly all of the bands are tabulated, along with the minimum uncertainties in the latter. Among the 414 bands, 135 (or 33%) were not reported in four previous, modern surveys of the DIBs in the spectra of various stars, including HD 183143. The principal result of this study is that the great majority of the bands in the catalog are very weak and fairly narrow. Typical equivalent widths amount to a few mÅ, and the band widths (FWHM) are most often near 0.7 Å. No preferred wavenumber spacings among the 414 bands are identified which could provide clues to the identities of the large molecules thought to cause the DIBs. At generally comparable detection limits in both spectra, the population of DIBs observed toward HD 183143 is systematically redder, broader, and stronger than that seen toward HD 204827 (Paper II). In addition, interstellar lines of $C_2$ molecules have not been detected toward HD 183143, while a very high value of $N(C_2)/E(B-V)$ is observed toward HD 204827. Therefore, either the abundances of the large molecules presumed to give rise to the DIBs, or the physical conditions in the absorbing clouds, or both, must differ significantly between the two cases.

## 1. INTRODUCTION

A new survey of the spectra of more than 30 OB stars with color excesses E(B-V) > 0.50 has been carried out at the Apache Point Observatory. These echelle spectra were recorded at a fairly high resolving power of R = 38,000 and at a high S/N ratio ≈ 1000 (per resolution

1 University of Chicago, Yerkes Observatory, Williams Bay, WI 53191
2 University of Chicago, Department of Astronomy and Astrophysics, 5640 South Ellis Avenue, Chicago, IL 60637
3 Also at Enrico Fermi Institute, University of Chicago, Chicago, IL 60637
4 University of Colorado, CASA-Campus Box 389, Boulder, CO 80309
5 Space Telescope Science Institute, 3700 San Martin Drive, Baltimore, MD 21218
6 University of Illinois, Departments of Chemistry and Astronomy, 600 South Mathews Avenue, Urbana, IL 61801
7 Embry-Riddle Aeronautical University, Department of Physics, 3700 Willow Creek Road, Prescott, AZ 86301
8 Johns Hopkins University, Department of Physics and Astronomy, 34th and Charles Streets, Baltimore, MD21218
9 University of Illinois, Department of Astronomy, 1002 W. Green St., Urbana, IL 61801
10 Current address: Department of Physics and Astronomy, Carthage College, 2001 Alford Park Drive, Kenosha, WI 53140, USA

element) near the H-alpha line. One of the purposes of this survey is to extend significantly the list of known diffuse interstellar bands (DIBs), principally by discovering new, generally narrow and weak examples. Owing to their possibly large numbers, such weak bands may be vital in eventually solving the longstanding challenge of convincingly identifying the various large molecules that are widely assumed to cause these absorption features (e.g. Herbig 1995;, Snow 1995, Hudgins & Allamonda 1999; Le Page et al 2003; Webster 2004; Sarre 2006; Cordiner & Sarre 2007; Pathak & Sarre 2008). Initial inspections of our various spectra indicated that two stars, HD 204827 and HD 183143, show especially notable differences among some of the DIBs present, despite the fact that both stars show similar, large values of interstellar reddening, E(B-V) = 1.11 and 1.27, respectively. Thus, the light paths toward these two stars may reveal DIB populations that lie toward opposite ends of the full range that is present in the solar neighborhood.

In Paper I of this series (Thorburn et al 2002), attention was primarily drawn to one of the most evident of these differences. In the spectrum of HD 204827, both the absorption lines of interstellar $C_2$ molecules and a particular subset of about 20 DIBs are much stronger than their counterparts in the spectrum of HD183143. We therefore referred to this subset as the $C_2$ DIBs. A well-defined statistical method was developed for identifying the relatively few bands that show this weak, secondary effect at a detectable level; the effect is easily masked by the much stronger dependence of DIB strengths on reddening. In Paper II (Hobbs et al 2008), a detailed account was given of the entire set of 380 DIBs found at $3900 < \lambda < 8100$ Å along the light path to HD 204827. The star is a double-lined spectroscopic binary (SB2; O9.5V + B0.5III), which apparently had not been previously observed specifically for DIBs. Thirty per cent of these 380 bands had not been detected in four previous, modern surveys of the DIBs seen toward various stars other than HD 204827.

In the present Paper III, the detailed, corresponding data will be presented for the bands seen in our spectrum of HD 183143, a star of appreciably later spectral type, B7Iae, than HD 204827. In contrast to the bands of HD 204827, the DIBs of HD 183143 have been extensively observed previously (e.g. Herbig 1975; Herbig & Soderblom 1982; Herbig 1988; Herbig & Leka 1991; Jenniskens & Desert 1994; Herbig 1995; Tuairisg et al 2000; Galazutdinov et al 2000b). Nevertheless, owing to the high quality of our spectra, it is likely that a significant number of new bands will be found. Furthermore, the detection limit of a few mA for the DIBs present in our final, combined spectrum of HD 183143 is generally similar, by design, to that for HD 204827. Thus, the two lists of DIBs can be usefully compared with minimal differences between the respective selection effects.

## 2. THE DATA

### 2.1. Observations and Data Reduction

The essential features of the overall observing program and of the reduction of the resulting spectra have been described previously in Papers I and II. Only a selective summary of those descriptions will be repeated here.

The spectra were obtained with the 3.5m telescope and the ARC echelle spectrograph (ARCES) at Apache Point Observatory. Each exposure provides nearly complete spectral coverage from about 3,700 to 10,000 Å at a resolving power R = 38,000 (Wang et al 2003).

The absolute sensitivity of the instrument (including the effects of atmospheric transmission) varies substantially over this wide wavelength range, reaching a broad maximum near 7100 Å for HD 183143. The star's spectrum was recorded on each of three nights in 1999: May 30, May 31, and June 11. Three untrailed, 30-minute exposures were obtained on May 30. These relatively long exposures of this bright, quite red star (V = 6.86, B-V = 1.22) were obtained in order to improve the relatively low S/N ratio otherwise achievable in the blue region. Each of these three exposures consequently is saturated at $\lambda > 6650$ Å and will not be used in any analyses of that wavelength region. Two 15-minute and six 5-minute exposures were obtained on May 31 and June 11, respectively, without any evidence of saturation. The 11 exposures obtained in a total observing time of 2.5 hours were combined during data reduction, with the exception that all results at $\lambda > 6650$ Å are based on only the latter eight exposures and 1.0 hr of observing time. The reduction of the individual exposures to a single, combined, final spectrum followed standard methods, except for several modifications needed for ARCES data (Paper I).

The primary procedural difference between our studies of HD 183143 and HD 204827 lies in the methods used to recognize, and to exclude, stellar lines. At least on the three nights spread over two weeks on which all of our observations of HD 183143 were acquired (Chentsov 2004), nearly all of the stellar lines showed invariant wavelengths. In contrast to those of the HD 204827 binary, the stationary stellar lines in our spectra of HD 183143 therefore cannot be separated from the DIBs on immediately conclusive, kinematic grounds. Instead, our spectrum of an unreddened, B8Ia comparison star, β Orionis, was used to carry out the identification of the stellar lines (Herbig 1975; Jenniskens & Desert 1994). Fortunately, the stellar spectra of the two stars are generally a good match, although some exceptions to this rule will be detailed in section 4.3. Even the widths (FWHM) of relatively weak lines, about 55 and 63 km $s^{-1}$ for β Ori and HD 183143, respectively, differ by only about 15%. The central absorption depths of many of the numerous stellar lines (and of many of the DIBs as well) amount to 1% or less of the stellar continuum, and the weaker lines of both kinds have not been previously catalogued for HD 183143 at the precision achieved here. For β Ori, Przybilla et al (2006) compiled a very convenient, partial list of stellar lines, and Chentsov & Sarkisyan (2007) presented a valuable spectral atlas with resolution, S/N ratio, and wavelength coverage very similar to those of our data here.

Observations were also obtained of several bright, broad-lined, early-type stars, to allow removal of telluric lines by division of the two types of spectra. At the level of precision just noted, this removal is generally not fully successful for the stronger telluric lines. The resulting, weak, residual telluric lines present can be readily recognized, as usual, by their fixed geocentric wavelengths from night to night, in the undivided spectra.

## 2.2 The S/N Ratio

Empirical estimates of the S/N ratio, per pixel, achieved as a function of wavelength in the final spectrum of HD 183143 are presented in column 2 of Table 1. The peak value, S/N ≈ 1000, is attained near 6400 Å, and the corresponding S/N ratios per resolution element are larger by a factor of $\sqrt{2.3} = 1.5$. On the coarse grid of wavelengths given in Table 1, the exclusion at $\lambda > 6650$ A of the exposures obtained on May 30 does not introduce an obvious step in the interpolated function. At all wavelengths, the S/N ratio for β Ori exceeds that for HD 183143. The disparities are generally fractionally small, but they become larger at $\lambda < 4900$ Å.

Three independent methods were used to determine or to corroborate the S/N values given in Table 1.

First, the S/N ratio in the continuum was measured directly in various wavelength segments of the spectrum which appear to be as free as possible from all types of absorption lines (interstellar, stellar, and residual telluric), to an absorption depth of about 1% of the continuum. Very few wavelength segments of this kind can be found. As a function of wavelength, the results of this process were fit by a quadratic function, but the scatter of the various data points about this function was unacceptably large. Therefore, the average spectrum from any one night was divided by its counterparts from the two other nights (or one other night, at $\lambda > 6650$ Å). The DIBs and the weaker stellar lines are generally removed from these quotients, to a typical precision of ± 0.3% of the continuum. As deduced from these quotients, the noise contributed by each of the respective, nightly averages to the full, final spectrum was then taken into account. The scatter about a quadrataic fitting function was significantly reduced by this procedure.

The second method began with the choice of 48 unblended, relatively narrow DIBs of intermediate strength. Representative parameters of these DIBs include a band width at half depth of FWHM = 0.72 Å, an equivalent width $W_\lambda$ = 13 mÅ, and a central absorption depth of 1.7% of the continuum. For each band chosen, the equivalent width was measured separately in each of the three averaged spectra respectively obtained on the various nights of our observations, and the resulting mean for all three nights and the corresponding rms deviation were then calculated. (At $\lambda > 6650$ Å, unsaturated spectra were available from only two nights, as noted above.) This rms error, $\Delta W_\lambda$, was converted to an effective S/N ratio by the formula

$$\Delta W_\lambda = 1.064 \times \text{FWHM} / (S/N), \qquad (1)$$

which expresses the equivalent width of a line of fractional central depth $\sigma = 1/(S/N)$ and of Gaussian profile. Many DIBs have irregular, asymmetric profiles, but most of those used for this particular purpose indeed are of crudely Gaussian form. As a function of wavelength, and after fitting by a quadratic function, the S/N ratio deduced from these DIBs showed acceptable external agreement with the results previously determined directly from the continuum, as well as smaller internal scatter about a quadratic fitting function. Therefore, the entries in column 2 of Table 1 were interpolated from the fitting function calculated from this second method alone. The uncertainty in these tabulated values is typically ± 15 %.

Finally, from the measured exposure levels cumulatively achieved in the combined, final spectrum of HD 183143, the S/N ratio expected theoretically from the photon shot noise alone was calculated. As a function of wavelength, the resulting function shows an amplitude and a shape quite different from those exhibited by the empirical results given in Table 1. Reached near 7100 Å, the maximum S/N ratio set by photon noise alone proves to be about 2150, a value well in excess of the maximum S/N ratio actually achieved. As a further test, the second method discussed above was extended to broader DIBs. To within the uncertainties, the resulting, empirical detection limit, $\Delta W_\lambda$, proves to be almost linearly proportional to FWHM (a result already used above), rather than to the square root of FWHM as would be expected if photon noise primarily limits the S/N ratio. These two facts together indicate that the errors arising in flat fielding, continuum placement, recognition of weak blending lines, and other systematic effects primarily establish the detection limits encountered in our final spectrum.

Although increasingly important at $\lambda < 4900$ Å and for the narrower DIBs, for which many of the systematic errors are minimized, the effects of photon noise are of secondary importance here for most of the DIBs.

The formula for $\Delta W_\lambda$ used above provides a practical measure of the minimum equivalent width that is detectable in our final spectrum. An overview of the resulting $1\sigma$ detection limits, $\Delta W_\lambda$, expected as a function of wavelength is presented in columns 3 and 4 of Table 1, for two illustrative values of the band width.

## 3.DATA ANALYSIS AND RESULTS

### 3.1 Background and Primary Results

As for HD 204837 in Paper II, our primary goal here is to compile a list of the DIBs that are present in the spectrum of HD 183143 above a detection limit near an equivalent width of a few mÅ. Our aim is to exclude from this list all features that are not genuine DIBs, while not being so restrictive as to exclude at the same time a significant number of very weak DIBs that are actually detectable in our spectra. The method used consists of two steps. First, the respectively averaged spectra from each of the three nights were intercompared by eye, in an attempt to identify all genuine DIBs that are detectable in our spectra. The primary attribute required of the DIB candidates is a satisfactory repeatability of the band's apparent strength and profile on all three nights of observation. The S/N ratios of the respectively combined spectra from the three nights are high enough in all cases so that this visual comparison could be carried out fairly simply and directly. A representative region of the spectra is shown in Figure 1; all wavelengths reported in this paper are those in air. Any important blends with stellar lines, residual telluric lines, or other DIBs were also noted. Both the spectrum of β Ori and, in many cases, the width and/or the shape of a feature in the spectrum of HD 183143 were used to exclude the stellar lines. The second step consisted of measuring the central wavelength, the band width (FWHM), and the equivalent width as well as its uncertainty for each of these bands, in the combined, final spectrum from all three nights. In a few cases, these subsequent measurements led to the rejection of a DIB candidate previously allowed by eye.

The results of these measurements are listed in Table 2 for 414 DIBs in the wavelength range $3900 < \lambda < 8100$ Å, and the combined, final spectrum of HD 183143 is displayed in Appendix A[9]. For completeness, four DIBs which are clearly present at $8100 < \lambda < 8774$ Å are also listed in Table 2, but the detection limits often are significantly poorer in this region than at the shorter wavelengths. Therefore, these four DIBs will not be included in the discussion or in the analyses here, and we will hereafter ignore their presence in Table 2. For each of the DIBs, columns 5 and 6 of Table 2 give the central wavelength and the band width (FWHM), respectively; columns 7 and 8 give the equivalent width and its uncertainty, respectively; and column 9 indicates the presence of significant blended features or other information. These various measurements will be described in more detail below. Columns 1 through 4 of Table 2 present the central wavelengths, as reported by the previous authors, of those DIBs detected in our spectra that had already been noted in one or more of the DIB atlases compiled by Jenniskens & Desert (1994), Galazutdinov et al (2000b), Tuairisg et al (2000), or Weselak et al (2000). The first two digits of those previous wavelengths are redundant with those in column 5 and have been suppressed. Occasional differences between the entries in columns 1 through 4 and their counterparts in Paper II reflect corrections of minor typographical errors (e.g. see 5986.57 Å), oversights (e.g. 5487.69 and 6468.75 Å), or other previous errors (e.g. 6788.84 Å). Column 13 provides sequential numbers for the DIBs, for convenient reference. For easy comparison, the format of columns 1 through 9 and column 13 are identical to those of the corresponding columns in Table 2 of Paper II.

[9] These tabular and graphical data are publicly available at http://dibdata.org.

The data in columns 11 and 12 partially characterize a stellar line that is blended with a DIB. These two columns respectively list the laboratory wavelength and the equivalent width of the stellar line, as measured in the spectrum of β Ori, *not* HD 183143. If more than one (isolated) stellar line interferes, typically for a broad DIB, the notation "> 1" appears in column 11, and the corresponding sum of the equivalent widths is given in column 12. When the notation "bl" is added to the entry in column 12, it signifies that the stellar line measured is itself a blend of at least two such lines. Finally, column 10 gives the central wavelength of the DIB as shifted to the rest frame of HD 183143, in order to allow easy comparison with the wavelength of the stellar line given in column 11. (The two leading digits of both wavelengths have also been suppressed.) The shift between the DIB wavelengths in columns 5 and 10 corresponds to the difference of 20.4 km $s^{-1}$ between the zero-points of the two wavelength scales (section 3.2). Under the assumption that the stellar line is identical in the spectra of β Ori and HD 183143, an approximate but useful picture of the blend in HD 183143 of the DIB and the stellar line is available by comparing the wavelengths and the equivalent widths presented in columns 7, 10, 11, and 12.

Except for the work of Weselak et al (2000), which excluded HD 183143, three of the four previous studies presented a list of DIBs that was derived from observations of HD 183143 along with several other stars. When our results for HD 183143 alone are compared with any that are based instead on averages over several stars, allowance must be made for at least two effects. First, the relative strengths of some DIBs vary significantly from one line of sight to another. Second, when the spectral types of two background stars differ appreciably, the blending with stellar lines may affect very differently the detectability of a given DIB in the spectra of the two stars. Thus, apparent discrepancies arising in such comparisons are not necessarily significant.

### 3.2. Central Wavelengths and Band Widths

The two components of the interstellar K I line recorded in our spectra (Figure 2) show respective heliocentric radial velocities of – 9.9 ± 0.4 and + 5.0 ± 0.6 km $s^{-1}$, or LSR velocities of + 8.6 ± 0.4 and + 23.5 ± 0.6 km $s^{-1}$. The zero-point of the DIB wavelength scale adopted here is set by assigning the laboratory wavelength of 7698.9645 Å (Morton 2003) to the stronger of the two K I components. We will refer to this as "the interstellar" wavelength scale and will measure DIB wavelengths on this system. When derived from unblended, generally weak, stellar lines, the heliocentric and LSR velocities of HD 183143 are + 10.5 ± 1.5 and + 29.0 ± 1.5 km $s^{-1}$, respectively, on each of our three nights. The other interstellar lines shown in Figure 2 will be discussed in section 5.3.

The laboratory wavelength could have been assigned equally well to the other component of the K I line, for example. This choice would change all of the absolute DIB wavelengths reported here by a shift that corresponds to the K I component splitting of 14.9 km $s^{-1}$, i.e. by 0.30 A at 6000 Å, the mid-point of the wavelength range analyzed here. This systematic shift is much larger than the random errors that arise in measuring the line-center wavelengths of most DIBs in our catalog. Depending upon such (currently arbitrary) choices made for wavelength zero-points, significant systematic differences are consequently expected among the absolute wavelengths reported for DIBs in various investigations. For this reason, specifically systematic differences among our wavelengths and the varied values found in columns 1 through 4 of Table 2 may be of little physical significance, in some cases.

In addition, precise definitions of the central wavelength and of the band width are necessary, in the common case of asymmetric DIBs. The definitions used here are those of Paper II, as follows. Within a DIB's profile, the midpoint along the vertical line drawn at fixed wavelength between the continuum and the point of deepest absorption is located. The horizontal line drawn through that midpoint intersects the shortest-wavelength and the longest-wavelength wings of the profile at wavelengths denoted by $\lambda_1$ and $\lambda_2$, respectively. Then, $\lambda_c = (\lambda_1 + \lambda_2) / 2$ and FWHM = $\lambda_2 - \lambda_1$. For example, the observed, overall width of the interstellar λ7699 K I line consequently amounts to FWHM = 0.65 Å, or 25.3 km $s^{-1}$. For comparison, the same line shows only a single component toward HD 204827 at ARCES resolution (Paper II), for which FWHM = 14.7 km $s^{-1}$. The appreciably larger velocity dispersion of the principal interstellar clouds toward HD 183143 is evident.

The smallest random errors of measurement that are attained for both $\lambda_c$ and FWHM typically amount to ± 0.04 Å for the narrower DIBs in Table 2, with widths near 0.5 Å. This uncertainty increases roughly in proportion to FWHM for the broader bands. The values of FWHM reported here are the measured values without any correction for instrumental broadening. The FWHM of the ARCES instrumental profile is 0.16 Å at 6000 Å, for example, while the narrowest DIBs in Table 2 show measured widths near 0.5 Å. The intrinsic widths of even the narrower DIBs must therefore be only marginally smaller than the measured values given in Table 2. For example, after quadratic subtraction of the instrumental width from the observed FWHM of 0.44 ± 0.04 Å, the intrinsic FWHM of the DIB at 5470.85 Å is about 0.41 ± 0.04 Å. This difference lies within the measurement errors.

In the current absence of identified DIB carriers and therefore of their laboratory spectra, no reliable method is available for deciding whether an irregular DIB profile with two (or more) distinct absorption maxima reveals one intrinsically asymmetric band or a blend of two narrower, more symmetric bands (e.g. Jenniskens et al 1996; Galazutdinov et al 2008; Kazmierczak et ak 2009)). The effectively arbitrary choices made here in such cases can be ascertained only from the numerical results reported in Table 2. For example, the profile of the band at 6376.21 Å is shown in Figure 3, and Table 2 shows that this asymmetric feature has been treated as a single DIB in our compilation. In two of the previous surveys cited in the table, this DIB was instead regarded as a blend of two distinct DIBs, a clearly plausible choice. To emphasize this difference in method, the notation "blend" has been entered in column 2 in place of the pair of distinct wavelengths reported by Galazutdinov et al (2000b). The corresponding comments apply to column 3. The notation "blend" is used throughout Table 2 to indicate that a feature treated here as one, asymmetric DIB was regarded in the previous investigation in question as two (or more) distinct, blended DIBs. This correspondence is sometimes ambiguous, and the set of "blend" entries indicated in Table 2 may occasionally be incomplete. If needed, the pertinent wavelengths of the separate DIBs are available in the previous papers.

3.3. Equivalent Widths and Their Errors

The equivalent widths presented in column 7 of Table 2 were calculated by integration over the DIB profiles in the combined, final spectrum formed from the 11 individual spectra obtained on all three nights (or eight spectra on two nights, at $\lambda > 6650$ Å). The corresponding $1\sigma$ errors listed in column 8 are estimates of the *minimum* errors. These errors were calculated from the formula $\Delta W_{\lambda} = 1.064 \times FWHM / (S/N)$ of section 2, using the values of the FWHM listed in Table 2 and the values of the continuum S/N ratio interpolated from Table 1.

The actual uncertainties in the equivalent widths must sometimes exceed the minimum values calculated routinely via the formula. The principal known source of additional uncertainty consists of blends with residual telluric lines, stellar lines, or other DIBs. A warning is provided in column 9 of Table 2 when such a blend significantly increases the uncertainties in measuring the wavelength, the width, and/or the equivalent width. A flag “d”, “s”, “t”, or “det” in column 9 identifies a blend with another DIB, a stellar line, a telluric line, or a detector artifact, respectively. The flags “s1” and “s2” indicate a blend specifically with a stellar N I or N II line, respectively; the role of the nitrogen lines will be detailed in section 4.3. In all of the flagged cases, an approximate deblending, by means of either Gaussian fitting or division by the spectrum of β Ori, has been carried out, if a definite value of the equivalent width has been specified for the DIB. If deblending instead appeared infeasible, only an upper limit is reported, although the DIB is definitely present. If the blending, usually with a strong stellar line, is so severe as to prevent useful measurement entirely, the notation “p” is entered in column 7 to indicate that the DIB is present. That is, we consider that all of the bands listed in Table 2 are definitely present in the spectrum of HD 183143. Among the 414 DIBs in Table 2 at $\lambda < 8100$ Å, all but four (at 4699.21, 5643.65, 6225.05, and 6831.21 Å) constitute detections at a statistical level of at least $3.2\sigma$, as based upon the minimum errors given in column 8.

Each of the well known DIBs at λ5780 and λ5797 shows a relatively deep, narrow absorption core superimposed on a much shallower, broader absorption feature with a complex, composite profile (Krelowski & Walker 1987; Jenniskens & Desert 1993). Owing in part to the difficulty in reliably defining the continuum for the broader features, the data for these two bands given in Table 2 refer to the respective narrow cores alone, as indicated by the “nc” flag listed in column 9

For the substantial number of very broad DIBs exemplified by the case of λ4428, an additional error enters into the data reduction. The wings of such broad bands extend over several orders of the echelle spectrum, and the process of correctly locating the continuum within this extended profile then becomes much more difficult. In such cases, the continuum level was interpolated smoothly across the affected gap between the many other, undistorted orders (Paper I). The minimum errors listed in Table 2 for the equivalent widths of the broad DIBs may be substantially increased by this additional uncertainty.

Preliminary values of the central wavelengths, the widths, and the equivalent widths were reported in Paper I (in Tables 2 and 3) for 32 of the bands listed in Table 2; the independent results presented here supersede the earlier ones. The definitions of $\lambda_c$ and FWHM described in section 3.2 were not applied in the earlier measurements, and some blends have been treated differently in the two cases. For example, the DIB at 6203.06 Å in Paper I corresponds approximately to the combination of bands in Table 2 at 6203.14 and 6205.20 Å, and the equivalent width of the λ5797.20 DIB was measured over a wider wavelength interval in Paper I. Except in cases of such procedural differences, the two independent measurements of the same data generally agree satisfactorily.

3.4 Selection Effects

The most sensitive detection limits are attained in our spectra within the approximate range 5100 < λ < 6865 Å. At shorter wavelengths, the poorer detection limits are attributable primarily to interference from many stellar lines and to progressively increasing photon noise. At longer wavelengths, the poorer sensitivity stems principally from three distinct difficulties: the many strong telluric lines present in much of the region; the progressively stronger interference fringes arising in the CCD detector, which degrade the accuracy of the flat-fielding and of the continuum definition; and the steadily increasing photon noise at λ > 7100 Å. In order to preserve a crudely uniform fractional completeness of detection, we will limit our subsequent analyses to the region 3900 < λ < 8100 Å. Nevertheless, the list in Table 2 undoubtedly is preferentially incomplete in the regions λ < 5100 Å and λ > 6865 Å. Even when these observational selection effects introduced by the various impediments are taken into account, it is nevertheless clear that the intrinsic spectral density of DIBs detectable toward HD 183143 decreases fairly sharply at λ < 4900 Å and at λ > 7800 Å.

Our survey is also biased against detection of broad DIBs and will be systematically incomplete in this way as well. At any fixed equivalent width, a relatively broad band must also show relatively shallow absorption, on average, because instrumental broadening is effectively negligible for all of the bands in Table 2. Both of these effects increase the difficulty of correctly interpolating the continuum across the DIB profile, so that the resulting systematic errors in determining the wavelength, width, and equivalent width of a band increase nearly linearly with increasing FWHM (Table 1). Therefore, at a given equivalent width, the fractional error in the equivalent width will also increase correspondingly, reducing the likelihood that progressively broader bands will be detected. In practice, this effect will be very large for the broadest bands. For example, the minimum 1σ detection limit for the λ4428 DIB, with FWHM = 22.5 Å, is 43 mÅ, larger by almost two orders of magnitude than the smallest uncertainties listed in Table 2 for narrow bands.

This conclusion is illustrated concretely by comparisons of our results with those of some of the previous studies of HD 183143. A perusal of Table 2 shows that 33% of the bands listed there were not detected in any of the four previous, reference surveys (section 5.1). Conversely, each of the four reference surveys lists definite detections of DIBs not found in our spectra. In comparisons of both kinds, the additional limitations imposed by observations of different stars must be taken into account (section 3.1) The results of Jenniskens & Desert (1994) and of Tuairisg et al (2000) are most useful for the second kind of comparison, because those authors also report the widths of the observed bands. A majority of the relatively small number of DIBs definitely observed by those authors but not detected by us are indeed broad, with FWHM > 6 Å. In contrast, only 9 of the 414 DIBs at λ < 8100 Å in Table 2 show such widths. That is, among the small fraction of DIBs that apparently have escaped our detection, most are certainly broad.

A particularly clear comparison of the results at λ < 8100 Å in Table 2 is possible with those of Herbig & Leka (1991) and Herbig (1995). Like ours, the latter results are based on data for HD 183143 alone, and estimates of the widths of the observed bands are similarly included. The agreement between the two sets of detected DIBs is generally quite good, after differences in the respective detection limits and in treating some blended DIBs are taken into account. Two principal differences emerge from the comparisons. (1) Because the detection limits for weak bands were less sensitive in the two earlier studies by a factor of roughly four, we have detected more than three times as many bands at $\lambda$ < 8100 Å. (2) All 14 bands reported in the earlier work that do not appear in Table 2 show FWHM > 8 Å. Some examples of these broad bands include the DIBs at 4060, 4180, 6177, 6940, 7429, and 7927 Å, all of which show FWHM > 13 Å. The incompleteness of Table 2 at FWHM > 6 Å is evident.

## 4.THE STELLAR LINES: SOME DETAILS

### 4.1. Deblending Procedures

In order to allow determinations of the parameters of the DIBs, either of two deblending techniques was applied to bands that are blended with stellar lines. For fairly symmetric DIBs that are either much narrower or much broader than a blended stellar line (or lines), a direct deblending into two or more Gaussian components was carried out. This method has the advantage that the spectrum of HD 183143 alone is used during the deblending. However, the results of this process may be insufficiently unique when both the central wavelengths and the widths of the two blended features are very similar. In addition, direct deblending into Gaussian components is usually inapplicable to the case of a noticeably asymmetric DIB. In these two circumstances, the stellar line was approximately removed by dividing the spectrum of HD 183143 by that of $\beta$ Ori. For identical equivalent widths, a stellar profile is narrower and deeper by about 15% in the spectrum of $\beta$ Ori, but the resulting errors introduced in determining the parameters of the DIBs are sometimes smaller than those encountered in direct deblending instead.

### 4.2.Possible Additional DIBs

The primary basis for recognizing and excluding stellar lines is their corresponding presence in the spectrum of $\beta$ Ori, an unreddened comparison star whose spectrum usually is a sufficiently good match to the stellar lines of HD 183143 for this purpose. In addition, the shapes and/or the widths of some DIBs can be used to recognize them as such. If it shows an equivalent width in excess of about 8 mA and is not appreciably blended with a stellar line, a DIB can often be recognized by an asymmetric profile alone. A comparison with $\beta$ Ori, if needed, can remove the possibility that it is a blend of two or more stellar lines. DIBs which are much broader or much narrower than the stellar lines can normally be reliably distinguished solely on that basis as well. Unsaturated stellar lines of HD 183143 show FWHM $\approx$ 63 km $s^{-1}$, while approximately half of the DIBs in Table 2 display FWHM $\leq$ 43 km $s^{-1}$ (in velocity units), in particular.

Recognizing and measuring a very weak DIB is normally more difficult, however, especially if it is blended with one or more stellar lines. The extreme example is provided by a DIB that (1) is blended with a stronger stellar line with little wavelength offset, and that (2) also displays a nearly symmetric, Gaussian band profile along with a band width that happens to be near that of the stellar lines. In this limit of a “perfect” blend, the comparison spectrum of $\beta$ Ori provides the sole test available. If such a DIB falls close to the detection limit, either measurement errors or a fractionally small, intrinsic difference between the strengths of the stellar line in the two stars can prevent proper assignment and measurement of the feature, as a DIB or as a stellar line. This conclusion contrasts very sharply with the case of HD 204827, in which the highly variable radial velocities of the stellar lines provided an additional, conclusive discriminant that is effectively independent of band strength. In short, the detection limit for weak bands is less objectively defined for HD 183143.

Therefore, Table 3 presents a list of 71 DIBs that are likely to be present in the spectrum of HD 183143, but with less certainty than those listed in Table 2. Except in the case of column 13, the formats of the two tables are identical. The assignment of a DIB to Table 3 rather than Table 2 requires a partially subjective decision. In addition, the list in Table 3 could be extended to include still further, possible bands, but it has been intentionally limited to what appear to be the more likely cases. No further use will be made in this paper of the bands listed in Table 3. Although broad (with FWHM > 4.4 Å) and, as a result, apparently strong, three of these possible DIBs remain uncertain primarily because they are also shallow. One of these three also lies at $\lambda$ > 8100 Å, a region not considered here in any detail.

Twenty-one of the 71 DIBs included in Table 3 have also been reported in at least one of the four earlier surveys and are therefore highly likely to be real; eleven of these 21 were reported in at least two of the previous surveys. Nevertheless, in order to preserve consistency of method insofar as possible, these 21 DIBs have been excluded from Table 2. Their inclusion would have increased the size of that sample by only about 5%.

4.3.Lines of N I and N II

Despite the decisively important, general agreement between the stellar lines of HD 183143 and β Ori, there are numerous, usually small differences between them. The most important difference is that all stellar lines of N I and N II present in the spectrum of β Ori are appreciably stronger in the spectrum of HD 183143, by factors in equivalent width that range from about 1.2 to 1.9. Some examples are shown in Figure 4. Because the sense of the disparity is the same for both stages of ionization that are accessible and for the full range of excitation potentials represented for each, the effect probably reveals a higher nitrogen abundance in HD 183143.

If unrecognized, such differences could introduce important errors into the present analyses. For example, a line of N I or N II that is marginally below the detection threshold in β Ori may be detectable in HD 183143. This feature could consequently be misclassified as a DIB, if the assumed abundance difference were unnoticed. The corresponding conclusion also applies, in principle, to strong lines such as those in Figure 4. As a result, we tabulated the nitrogen lines expected to be present in either spectrum, based on laboratory data and on the lines observed to be present in the two spectra. Fortunately, the resulting number of near coincidences between these stellar nitrogen lines and the observed DIBs proved to be small. The DIBs known to be affected by blends with stellar N I or N II lines are emphasized in Tables 2 and 3 by the special flags “s1” and “s2”, respectively. For these lines identified with nitrogen, the difference between the applicable laboratory wavelength and the wavelength measured in the spectrum of β Ori (column 11, Tables 2 and 3) was required to be less than ± 0.1 Å. When the deblending of a stellar nitrogen line was accomplished through division by the spectrum of β Ori, the reduced equivalent width of the stellar line in that spectrum will introduce additional errors into the derived parameters of the DIB. Most of the upper limits listed in column 7 of Table 2 arise in this way.

Some lines of other elements also show differences in equivalent width between HD 183143 and β Ori, although no other case seems to be as systematic and as marked as that of nitrogen. Some, but not all, lines of O I appear weaker in HD 183143 than in β Ori (Figure 4). This effect cannot cause stellar O I lines in HD 183143 to be misidentified as DIBs owing to an absent counterpart in β Ori, so this effect appears unimportant for our specific purposes. Some of the stellar Si II lines are also relatively weak in the spectrum of HD 183143.

### 4.4 Radial-Velocity Variability

Our observations of HD 183143 were confined to three nights within a period of two weeks in 1999. Unblended, weak lines of He I, Si II, S II, and Fe II yield a heliocentric radial velocity of +10.5 ± 1.5 km $s^{-1}$ that is constant for all of these lines on all three nights. Five lines of N I show unexplained blueshifts of up to 8 km $s^{-1}$ relative to these lines.

More generally, over a much longer interval, Chentsov (2004) found significant variations in the apparent radial velocity of HD 183143. On the basis of eight echelle spectra acquired during nine years beginning in 1993, he discovered complex, apparently aperiodic variability of the radial velocities derived from most stellar lines. The variability differs among the various atoms and ions observed. The radial velocities derived from weak lines ranged from 9 to 20 km $s^{-1}$ during this interval, and the value observed on 5 June 1999, +10 km $s^{-1}$, agrees with our nearly contemporaneous results reported here. Chentsov attributed the observed variability to non-stationary behavior in the atmosphere of this very luminous star and in the wind emanating from it; no evidence that the star is a binary was reported. Chentsov further classified HD 183143 as a rare, Galactic hypergiant, sometimes also referred to as a candidate Luminous Blue Variable (Humphreys & Davidson 1994; Stahl et al 2003). On that basis, the star lies at a distance of about 2 kpc, at Galactic coordinates (l,b) = (53.3°,+0.6°). It does not appear to be a member of a cluster or group.

## 5. DISCUSSION

### 5.1. New Bands

In the analysis in Paper II of the DIBs in the spectrum of HD 204827, “new” bands were defined as those that had not been reported in any of the four previous DIB surveys (toward stars other than HD 204827) cited in Table 2. In some cases, those DIBs may actually have been reported previously, elsewhere in the literature. Owing to the large number of previous investigations of the DIBs seen toward HD 183143 (section 1), these four particular surveys, by themselves, form a much less complete reference sample for the analysis here. Nevertheless, in the interests of continuity and simplicity, the same meaning will be understood here for “new” DIBs. This definition also excludes a comparison specifically with the DIBs tabulated for HD 204827 in Paper II.

In the region $3900 < \lambda < 8100$ Å, the total number of DIBs detected in the spectrum of HD 183143 is 414, of which 135, or 33%, are new by this limited definition (Table 2). (The possible bands listed in Table 3 will not be considered in any of our analyses.) Both the number of DIBs detected and the corresponding fraction of new bands are slightly higher than for HD 204827. In the region $3900 < \lambda < 5620$ Å, 30 of the 50 DIBs detected toward HD 183143, or 60%, did not appear in the four reference surveys, in a spectral region where the average spacing between adjacent pairs of these 50 features is 34.4 Å (Table 4). The corresponding values for the region $6190 < \lambda < 6450$ Å reveal a much lower discovery fraction amounting to 8 of 50, or 16%, in a region with a much smaller average spacing of 5.2 Å between adjacent DIBs. Thus, our discovery fraction is higher where the spectral density of DIBs is lower. Similar statistical results are collected in Table 4 for the full spectral range included in this study. The table is organized not by uniform wavelength intervals, but rather by successive groups of 50 DIBs each, in order of increasing wavelength.

A quantitative test shows that the confusion potentially introduced by overlapping DIBs appears to play a negligible role, however. For this purpose, the various wavenumber splittings between the 413 pairs of immediately adjacent DIBS at $\lambda < 8100$ Å in Table 2 are analyzed. Differences in wavenumber, rather than in wavelength, are used for this purpose, because the former are proportional to the corresponding energy differences, which are directly related to the presumed molecular structure. As a function of wavenumber splitting, the histogram of Figure 5 plots the logarithmic numbers of splittings found in ten bins. All of these bins are 10 $cm^{-1}$ wide, their centers are separated by 10 $cm^{-1}$, and the first of these bins is centered at 5 $cm^{-1}$. At $\lambda = 5000$ Å, an interval $\Delta\sigma = 10$ $cm^{-1}$ corresponds to $\Delta\lambda = 2.5$ Å. A cutoff in the histogram at large splittings has been introduced owing to the nearly vanishing populations of those bins; only 21 splittings are thereby excluded from the total of 413. The approximately linear relation seen in Figure 5 reveals that the number of splittings declines nearly exponentially with increasing splitting. This result is expected if the wavenumber differences are randomly distributed and if there is negligible overlap between adjacent members of this sample (Welty et al 1994; see their Figure 7). The central result for the present purpose is that no relative falloff in numbers is seen in the bins containing the smallest splittings. This histogram further suggests that the wavenumber splittings are distributed randomly, so that no preferred splitting established by molecular structure is evident in this particular test of the sample.

## 5.2. Observed Distributions of Wavelengths, Widths, and Equivalent Widths

The histogram of Figure 6 shows the distribution by wavelength of all 414 DIBs at $\lambda < 8100$ Å in Table 2. The first populated bin spans the range from 4300 to 4500 Å and contains only three bands. The bins containing the highest number of DIBs are centered at $6000 \leq \lambda \leq 6800$ Å. This high spectral density of DIBs may extend to still longer wavelengths, since strong telluric absorption almost certainly obscures a significant fraction of the weaker DIBs actually present at $\lambda > 6865$ Å. The previous, corresponding results for HD 204827 are also shown in Figure 6 for comparison. The spectral density of those DIBs reaches an intrinsic maximum near 6100 Å. The population of DIB wavelengths along the light path to HD 183143 clearly is systematically redder than that toward HD 204827.

Figure 7 shows the distribution by observed band width of the 413 DIBs in Table 2 at $\lambda < 8100$ Å for which definite values were measured. The bin at FWHM = 1.75 Å contains the 66 bands with FWHM > 1.70 Å, and the median width for the full sample is 0.91 Å. The first populated bin is centered at 0.45 Å, extends from 0.4 to 0.5 Å, and contains the 15 narrowest DIBs in the sample. The most highly populated bins are centered at widths of 0.65 and 0.75 Å. The falloff in numbers at FWHM < 0.65 Å must result from at least three effects: (1) instrumental broadening; (2) an unknown lower limit to the intrinsic widths of most DIBs, within a single interstellar cloud; and (3) to an unknown extent, the velocity separation of 14.9 km $s^{-1}$ (or 0.30 Å at 6000 A) between the two groups of interstellar clouds seen in Figure 2. Also shown in Figure 7 are the corresponding data for HD 204827. The median FWHM for that full sample of DIBs is 0.62 Å. The most highly populated bin toward HD 204827 is centered at 0.55 Å, and, relative to HD 183143, the smaller proportion of large band widths is especially evident in the bins centered at FWHM ≥ 0.65 Å. While this difference is probably caused in part by the smaller velocity dispersion of the interstellar clouds along that light path, all three effects noted above become progressively less important at these larger band widths. Thus, the population of DIB widths toward HD 183143 clearly is selectively broader than that toward HD 204827. The same conclusion follows when the band widths plotted as the abscissae in Figure 7 are the normalized values, FWHM/$\lambda$.

Figure 8 shows the distribution by equivalent width of the 401 DIBs in Table 2 at $\lambda <$ 8100 Å for which definite values were measured. The first bin extends from 0 to 2 mÅ and contains only seven bands, of which the two nominally weakest are DIBs at 5838.09 and 6831.21 Å, which show $W_\lambda = 1.9 \pm 0.5$ mÅ and $W_\lambda = 1.8 \pm 0.7$ mÅ, respectively. The sharp falloff in the distribution seen in the two bins centered at $W_\lambda \leq 3$ mÅ reflects the approximate detection limit for the narrower bands in our study. For the broader DIBs, this detection limit must occur at proportionally larger equivalent widths (Table 1). The most important overall result is the progressively decreasing number of DIBs at $W_\lambda \geq 5$ mÅ. When very weak DIBs can be efficiently detected, their numbers dominate the corresponding distribution. The corresponding data for HD 204827 are also shown in Figure 8. The most outstanding difference between the two distributions is the relatively large number of very weak bands that are seen toward HD 204827. This result presumably reflects primarily the difference between the respective distributions of band widths already noted. The selectively broader DIB population toward HD 183143 biases the corresponding detection limit to higher values, at comparable S/N ratios.

Along with the relative strengths of the $C_2$ DIBs, the contrast between the distributions of the DIB wavelengths and band widths in the spectra of HD 183143 and HD 204827 is notable. All three effects clearly point to a difference between the respective relative populations of the molecular carriers or to markedly different physical conditions in the two sets of clouds, or both. A more detailed comparison of the sets of DIBs seen toward these two stars will be given in a subsequent paper in this series.

### 5.3 The Narrowest DIBs

The respectively strongest interstellar lines of Ca I, CN, CH and $CH^+$ in the spectrum of HD 183143 are also displayed in Figure 2. Except for CN, these lines clearly reveal two components at nearly the same velocities as the much stronger K I lines. The relative strengths of the two components of the molecular lines are opposite to those of the atomic lines. Although their shapes become obviously asymmetric, neither component of the relatively strong K I, CH, or $CH^+$ lines is resolved into further, clearly separated components at the much higher resolving power of R = 200,000 near K I, or at R = 120,000 near CH and $CH^+$ (McCall et al 2002). We temporarily assume that the molecules thought to cause the DIBs are distributed nearly equally between the two cloud groups, as in the cases of K I and $CH^+$, and that the intrinsic band width of a typical DIB is not narrower than that of either K I line component. In that case, model profiles show that the combination of the velocity difference between the two K I line components (14.9 km $s^{-1}$), the intrinsic width of either K I component (FWHM ≈ 7 km $s^{-1}$, after correction for instrumental broadening), and the instrumental broadening (8 km $s^{-1}$) sets a lower limit of about 25 km $s^{-1}$ on the observed DIB band widths (section 3.2). Any DIB observed to be definitely narrower than the observed width of the K I lines may suggest that the molecules which cause the DIB are strongly concentrated into only one of the two cloud groups.

One representative group of the narrowest DIBs consists of the 15 bands that populate the bin centered at FWHM = 0.45 Å in Figure 7. The velocity widths of these bands are collected in Table 5, along with other pertinent data. All of these DIBs are also very weak, with $W_\lambda \leq 4.6$ mÅ; as a result, a typical fractional error in measuring these widths is perhaps 20%. Therefore, even the narrowest of the tabulated widths, 19.3 km $s^{-1}$, probably does not conclusively indicate a marked difference in the distribution of the pertinent molecules between the two cloud groups, as compared to K I, CH, and $CH^+$.

A somewhat deeper, stronger, but still narrow DIB generally shows a more accurately defined profile whose fractional width can be measured with higher precision. An illustrative example is the DIB at 7385.98 A, which shows $W_\lambda$ = 15.2 mÅ and FWHM = 0.64 Å, or 26.0 km $s^{--1}$ (Figure 9); other, similar examples are provided by the bands at 6223.66, 6553.95, and 7322.11 Å. Calculated blends of two weak, Gaussian line components of equal strength that are separated by 15 km $s^{-1}$ cannot satisfactorily duplicate the observed λ7386 profile. For intrinsic band widths in excess of about 13 km $s^{-1}$, the theoretical, instrumentally broadened profile is clearly wider than the observed one. At narrower intrinsic widths, the theoretical profile is markedly flat-bottomed or partially resolved into two components, and these shapes disagree with the observed profile. On the other hand, after instrumental broadening, a theoretical profile which consists of a single, nearly Gaussian component with an intrinsic FWHM = 24.7 km $s^{--1}$ can readily duplicate these observed λ7386 profile. The simplest hypothesis is that, like CN, the associated DIB molecule is substantially concentrated into only one observed cloud group or the other. If true, this result could apply more extensively, to other DIBs in Table 2.

This tentative conclusion is further supported by the properties of the λ7386 DIB observed in the spectrum of HD 204827, where the band is somewhat weaker and narrower but shows a generally similar profile (Paper II). In this case, the width of the DIB exceeds that of the corresponding interstellar K I line by about 40%, in marked contrast to the instance of HD 183143, where the two widths are effectively identical. At least for these two stars, the narrow width of this DIB becomes much less conspicuous when it is not confused by an array of absorbing clouds that spreads over a velocity range at least comparable to the DIB's intrinsic width. Observations at much higher resolution of the λ7386 band and similar DIBs in the spectra of both stars might help to illuminate this effect directly

To test a possible alternative explanation, we also carried out a comparison of the wavelengths in Table 5 with the relevant, known wavelengths of 23 diatomic, triatomic, or tetraatomic molecules, including 19 not yet detected by means of interstellar absorption lines. The search was unsuccessful in identifying a match of this kind. The 23 molecules considered are CH, $CH^+$, $CH^-$, NH, $NH^+$, $NH^-$, OH, $OH^+$, $OH^-$, NaH, MgH, $MgH^+$, AlH, SiH, $SiH^+$, SH, $SH^+$, CaH, FeH, $CH_2$, $CH_2^+$, $NH_2$, and $CH_3$.

## 5.4 A Further Test for Molecular Structure

The structures of the molecules that are presumed to give rise to the DIBs may introduce some recognizable signatures into the pattern of the wavelengths in our relatively large sample of DIBs. The histogram of Figure 5 provided no such evidence, but it was based on the wavenumber splittings among the nearest-neighbor DIBs alone. The much larger sample of wavenumber splittings available from all possible pairs of DIBs listed in Table 2 permits a further test (Herbig 1995).

Among the 414 bands at $\lambda < 8100$ Å, the total number of distinct pairs of DIBs available is 85,491. A conveniently smaller sample can be isolated by considering only those DIB pairs whose wavenumber splittings do not exceed some upper limit. As an example, an upper limit of 400 $cm^{-1}$ was chosen, which yields 11,334 DIB pairs. The corresponding wavenumber splittings are distributed as shown in Figure 10, where the splittings have been grouped into 40 bins. All of these bins are 10 $cm^{-1}$ wide, their centers are separated by 10 $cm^{-1}$, and the first of these bins is centered at 5 $cm^{-1}$. The observed distribution declines smoothly from an average of about 325 pairs per bin at the smaller splittings to about 220 pairs per bin at the larger splittings, with no statistically significant excesses or deficiencies in any of the individual bins. The uncertainty expected in the number of pairs in a bin is comparable to the square root of that number. (The intrinsic scatter in this distribution presumably is established by the absorbing molecules and probably does not assume a Gaussian form.) As in Figure 5, the pertinent molecular energy-level differences appear to be effectively distributed randomly in our sample, without any reliably identified, preferred values.

## 6. SUMMARY

The diffuse interstellar bands detectable in high-quality echelle spectra of HD 183143 obtained on three different nights are investigated here. The principal conclusions derived from these data are the following.

1. A total of 414 DIBs are detected toward HD 183143 in the range $3900 < \lambda < 8100$ Å. The primary DIB population consists of bands that are fairly narrow (FWHM near 0.7 Å), very shallow (< 1% fractional absorption), and therefore quite weak ($W_\lambda$ near a few mÅ). The central wavelengths, the widths, and the equivalent widths of nearly all of these bands are presented in Table 2, along with the minimum errors for the latter values.
2. Among these 414 DIBs, 135 (or 33%) were not reported in four previous, modern surveys of the DIBs in the spectra of various stars, including HD 183143. By themselves, these surveys constitute only a partial, although convenient, indicator of the extensive, previous studies of the DIBs in the spectrum of HD 183143. The principal observational selection effects in our results are described in section 3.4.
3. For most of the DIBs, the systematic zero-point uncertainty in all of the absolute wavelengths exceeds substantially the random errors of measurement, which amount to ± 0.04 Å for the narrowest bands. This systematic uncertainty arises from the unknown identities and laboratory spectra of the presumed absorbing molecules, and from the presence of multiple clouds along the light path. Two groups of clouds evident at the interstellar K I lines are separated by 14.9 km $s^{-1}$, or 0.38 Å at the λ7699 line.

4. The DIBs have been distinguished from stellar lines in the spectrum of HD 183143 primarily by a comparison with the spectrum of the unreddened star β Ori. With some exceptions, such as all lines of nitrogen, the two sets of stellar lines are fairly well matched. Even the line widths differ in the two spectra by only 15%. The widths and/or the sometimes-asymmetric profiles of the DIBs often differentiate them from stellar features in HD 183143 as well.

5. Over 200 Å ranges, the average spacing between adjacent DIBs decreases to a minimum of 2.8 Å near 6800 Å. However, the distribution of the wavenumber gaps between adjacent bands appears to be both effectively random and substantially unaffected by mutual overlapping at R = 38,000. In particular, among adjacent DIBs, no preferred wavenumber interval is evident that might be identified with the molecular structure of a specific DIB absorber. Similarly, no preferred wavenumber splitting is evident from a single, limited investigation of the wavenumber splittings among all possible pairs of DIBs in the catalog.

6. An additional list is presented in Table 3 of 71 DIBs which may possibly be present in our spectra of HD 183143. Most are very weak bands. At least 21 of these 71 were also reported in previous investigations and are highly likely to be real.

7. At generally comparable detection limits in both spectra, the population of DIBs observed toward HD 183143 is systematically redder, broader, and stronger than that seen toward HD 204827. In addition, interstellar lines of $C_2$ molecules have not yet been detected toward HD 183143, and the ratio $N(C_2)/E(B-V)$ observed toward HD 204827 exceeds that toward HD 183143 by a factor of at least 78 (Paper I). Therefore, either the relative abundances of the molecules presumed to give rise to the DIBs, or the physical conditions in the absorbing clouds, or both, must be significantly different in the two cases.

It is a pleasure to thank John Barentine, Jack Dembicky, William Ketzebach, Russet McMillan, and Gabrelle Saurage for consistent and diligent help. This research was funded partly by the University of Colorado and NASA. B.J.M .gratefully acknowledges support from the David and Lucille Packard Foundation and the University of Illinois. P.S. acknowledges support from a NASA Spitzer grant under RSA agreement 1314055. T.O. is supporated by NSF AST 0849577.

# APPENDIX A

A general overview of the final, combined spectrum of HD 183143 and of the spectrum of β Ori is provided in Figure 11, where the DIBs present in the spectrum of HD 183143 are marked by vertical lines. The lines are numbered in order of wavelength, at intervals of five bands, for easy comparison with the entries in Table 2. The wavelength scale in Figure 11 is insufficiently expanded to allow detailed scrutiny of most of the DIBs included, but their more important characteristics are listed in Table 2. The imperfect removal of telluric lines in Figure 11 is often obvious, especially in the region at $\lambda > 6865$ Å. Almost all of the apparent emission lines in both spectra in that region result from erroneous telluric corrections and are spurious. A few exceptions in the spectrum of HD 183143 are Hα, two lines of Mg II at 7877.05 and 7896.37 Å, and an unidentified line near 7849.3 Å, all of which show P Cygni profiles or analogous absorption. Paper I presented preliminary plots at a more expanded scale of our spectra of both HD 183143 and β Ori. The plots display fifteen selected wavelength regions, each 50 Å wide, located at $4350 < \lambda < 6750$ Å.

A number of atomic and (diatomic) molecular interstellar lines are also present in the spectrum of HD 183143 (Gredel et al 1993; Galazutdinov et al 2000a; McCall et al 2002). These lines definitely detected in our spectra are reported in Table 6 along with their strengths. The upper limits cited for the D lines of Na I reflect blends with relatively weak stellar lines. These stellar lines cannot be removed straightforwardly, owing to the obscuring interstellar Na I lines also present toward β Ori. The upper limits given for HD 183143 are the total equivalent widths for the Na I blends. The lines listed in Tables 6 are identified in Figure 11 by asterisks placed above the spectrum of β Ori, although the asterisks refer to the spectrum of HD 183143. The first seven lines listed in Table 6 lie outside the wavelength limits adopted for Table 2 and Figure 11. However, these narrow lines can be measured reliably in our final spectrum and are therefore included in Table 6 as well.

FIGURE CAPTIONS

Fig. 1. Spectra of β Ori (top plot) and of HD 183143 (bottom four plots) in the region near 6070 Å. For HD 183143, an average of the three respective nightly averages (bottom three plots) appears above those three. The various spectra are displaced vertically by 2% of the continuum. The wavelength scale plotted is that in the rest frame of the stronger of the two interstellar K I components seen toward HD 183143 (section 3.2), and the wavelength scale for β Ori has been arbitrarily shifted so as to align the stellar lines in the two spectra. A stellar Ne I line with a laboratory wavelength of 6074.338 Å is seen in all five spectra. Seven relatively narrow DIBs appear only in the spectra of HD 183143, at wavelengths of 6057.61, 6059.41, 6060.38, 6065.31, 6068.39, 6071.22, and 6081.19 Å. In addition, relatively broad bands appear at 6065.98 (Table 2) and, possibly, 6079.11 Å (Table 3). The feature at 6068.39 Å could alternatively be characterized as a blended pair of distinct, quite narrow bands, but it is treated here as a single, asymmetric DIB. The equivalent width of the band at 6081.19 Å is 2.8 ± 0.5 m Å.

Fig. 2. Interstellar lines of Ca I, CN, CH, $CH^+$, and K I (from top to bottom) in the spectrum of HD 183143 at a resolution of 8 km $s^{-1}$. The respective rest wavelengths are 4226.73, 3874.60, 4300.32, 4232.54, and 7698.96 Å. The spectra are shown on a heliocentric velocity scale, and adjacent plots are separated vertically by 12% of the continuum. "The interstellar" wavelength scale specified here is that in the rest frame of the stronger of the two K I line components, which show heliocentric radial velocities of -9.9 and +5.0 km $s^{-1}$. On the three nights of our observations, the heliocentric radial velocity of HD 183143 was +10.5 km $s^{-1}$, and the width of weak, unblended stellar lines was FWHM = 62 km $s^{-1}$. The interstellar $CH^+$ line lies in the blue wing of the stellar Fe II λ4233.17 line, while the interstellar CH line lies in the red wing of a very weak stellar line. In the case of CN, the weak feature near –13.4 km $s^{-1}$ may consist of noise only, and the feature near –41.3 km $s^{-1}$ is the long-wavelength component of the adjacent CN feature at a rest wavelength of 3874.00 Å (Appendix A, Table 6). The relative strengths of the two line components are reversed between the atomic and the molecular lines.

Fig. 3. The spectra of HD 183143 (below) and β Ori (above) in the region of the DIBs at 6376.21 and 6379.32 Å in HD 183143. The interstellar wavelength scale is shown, and the comparison spectrum has been shifted upward by 2% of the continuum. Fairly evident in this spectrum is a weak, unidentified stellar line, which in HD 183143 is marginally seen as a blend with the long-wavelength wing of the λ6379 DIB.

Fig. 4. The spectra of HD 183143 (below) and β Ori (above) in the region of three stellar N II lines (at 4601.48, 4607.16, and 4613.87 Å) and a blend of two stellar O II lines (near 4596.0 Å). In contrast to Figures 1 and 3, the wavelength scale is that in the rest frame of the stars. The comparison spectrum has been shifted upward by 1% of the continuum. The three N II lines are slightly broader and evidently deeper in HD 183143; their equivalent widths exceed the corresponding values in β Ori by an average of 60%. In contrast, the O II line is weaker in HD 183143 by about 19%.

Fig. 5. The distribution, by wavenumber splitting, of the separations between 392 pairs of adjacent DIBs at $\lambda < 8100$ Å in the spectrum of HD 183143 (Table 2); 21 additional, large splittings, $\Delta\sigma > 100$ $cm^{-1}$, are not shown. The vertical scale gives the logarithms of the numbers of pairs in each of ten bins. All bins are 10 $cm^{-1}$ wide, their centers are also separated by 10 $cm^{-}$

$^{1}$, and the first bin is centered at 5 $cm^{-1}$. At $\lambda$ = 5000 Å, an interval $\Delta\sigma$ = 10 $cm^{-1}$ corresponds to $\Delta\lambda$ = 2.5 Å.

Fig. 6. The distribution by wavelength of 414 DIBs at $\lambda$ < 8100 Å in the spectrum of HD 183143 (filled circles), from Table 2. The wavelength bins are 200 Å wide, their centers are also spaced by 200 Å, and the first bin is centered at 4000 Å. The corresponding data for 380 bands in the spectrum of HD 204827 (open squares) are also shown (Paper II).

Fig. 7. The distribution by band width of the 413 DIBs at $\lambda$ < 8100 Å in the spectrum of HD 183143 (filled circles) for which definite values were measured (Table 2). The widths are the measured values uncorrected for instrumental broadening, which amounts to 0.16 Å at 6000 Å. The bins are 0.1 Å wide, their centers are also spaced by 0.1 Å, and the first bin is centered at 0.05 Å. The bin at FWHM = 1.75 Å accounts for the 66 DIBs with FWHM ≥ 1.70 Å. The corresponding data for 376 bands in the spectrum of HD 204827 (open squares) are also shown (Paper II).

Fig. 8. The distribution by equivalent width of 401 DIBs at $\lambda$ < 8100 Å in the spectrum of HD 183143 (filled circles) for which definite values were measured (Table 2). The bins are 2 mÅ wide, their centers are also spaced by 2 mÅ, and the first bin is centered at 1 mÅ. The bin centered at $W_\lambda$ = 41 mÅ accounts for the 63 bands with $W_\lambda \geq$ 40 mÅ. The corresponding data for 360 bands in the spectrum of HD 204827 (open squares) are also shown (Paper II).

Fig. 9. The spectra of HD 183143 (below) and β Ori (above) in the region of two relatively narrow DIBs. The interstellar wavelength scale is plotted, and the two spectra are displaced vertically by 1% of the continuum. An unidentified, very weak stellar line appears to be present in β Ori, near the center of the wavelength range shown.

Fig. 10. The distribution of the wavenumber splittings between 11,334 pairs of DIBs with separations not exceeding 400 $cm^{-1}$. The bins are 10 $cm^{-1}$ wide, their centers are also spaced by 10 $cm^{-1}$, and the first bin is centered at 5 $cm^{-1}$. The error bars span $\pm\sqrt{N}$, where N is the number of splittings in each bin.

Fig. 11. Final, combined spectrum of HD 183143 in the range 3900 < $\lambda$ < 8100 Å, along with the spectrum of β Ori. See the description in Appendix A. The upper and lower panels on each page display the same data at different vertical scales; both scales are appreciably expanded in order to emphasize the generally weak DIBs. The DIBs present in the spectrum of HD 183143 are marked by vertical lines and are numbered in order of increasing wavelength. The atomic and diatomic interstellar lines listed in Table 6 are identified by asterisks placed above the spectrum of β Ori (in order to avoid crowding), although the asterisks refer to the spectrum of HD 183143. Only three illustrative pages among the total of 42 are printed here; the entire figure is publicly available at http://dibdata.org.

Table 1

1σ Detection Limits [a]

| $\lambda$(Å) | S/N[b] | $\Delta W_\lambda$ (mÅ) FWHM = 0.5 Å | $\Delta W_\lambda$ (mÅ) FWHM = 2.0 Å |
|---|---|---|---|
| 4000 | 327 | 1.6 | 6.5 |
| 4500 | 588 | 0.9 | 3.6 |
| 5000 | 786 | 0.7 | 2.7 |
| 5500 | 922 | 0.6 | 2.3 |
| 6000 | 994 | 0.5 | 2.1 |
| 6500 | 1004 | 0.5 | 2.1 |
| 7000 | 971 | 0.5 | 2.2 |
| 7500 | 953 | 0.6 | 2.2 |
| 8000 | 935 | 0.6 | 2.3 |

[a] A significant fraction of the region $\lambda > 6860$ Å is effectively blocked by strong telluric absorption lines.

[b] Per pixel.

Table 2

Diffuse Interstellar Bands in the Spectrum of HD 183143

| JD94 $\lambda_c$(Å) | GM00 $\lambda_c$(Å) | TC00 $\lambda_c$(Å) | WS00 $\lambda_c$(Å) | $\lambda_c$(Å) | FWHM (Å) | $W_\lambda$ (mÅ) | $\Delta W_\lambda$ (mÅ) | note | DIB $\lambda_c$*(Å) | β Ori $\lambda_c$*(Å) | β Ori $W_\lambda$ (mÅ) | # |
|---|---|---|---|---|---|---|---|---|---|---|---|---|
| | | | | 4371.73 | 1.03 | 10.8 | 2.1 | s | 71.43 | 71.26 | 6.3 | 1 |
| 28.88 | | 27.96 | | 4428.83 | 22.56 | 5700.0 | 43.3 | s | 28.52 | > 1 | -- | 2 |
| | | | | 4494.55 | 2.09 | 20.1 | 3.8 | s | 94.24 | 93.28 | 2.3 | 3 |
| 1.80 | 1.80 | 1.65 | | 4501.66 | 3.01 | 211.2 | 5.4 | s | 1.35 | 1.40 | 7.1 | 4 |
| | | | | 4650.77 | 1.61 | 17.3 | 2.6 | s | 50.45 | 50.79 | 5.1 | 5 |
| | | | | 4699.21 | 1.36 | 5.0 | 2.1 | s | 98.89 | 99.16 | 1.9 | 6 |
| blend | 26.27 | blend | | 4727.16 | 3.11 | 156.2 | 4.8 | s | 26.84 | 26.74 | 11.1 bl | 7 |
| 62.57 | 62.67 | 62.57 | | 4762.62 | 2.50 | 126.5 | 3.8 | s | 62.30 | > 1 | 8.1 | 8 |
| 80.09 | 80.04 | 80.10 | | 4780.24 | 1.72 | 68.1 | 2.6 | s | 79.92 | 79.38 | 15.6 bl | 9 |
| 81.83 | | 82.56 | | 4881.06 | 11.31 | 343.7 | 16.2 | s | 80.73 | > 1 | 49.8 | 10 |
| | | | | 4957.02 | 1.42 | 9.7 | 2.0 | | 56.68 | | | 11 |
| 63.96 | 63.90 | 63.89 | | 4963.98 | 0.72 | 26.4 | 1.0 | | 63.64 | | | 12 |
| | | | | 4969.27 | 0.89 | 7.5 | 1.2 | s | 68.93 | 68.13 | 13.6 bl | 13 |
| 84.73 | 84.81 | 84.78 | | 4984.87 | 0.61 | 11.5 | 0.8 | s | 84.53 | 84.36 | 2.7 | 14 |
| | | | | 4987.68 | 1.23 | < 9.6 | 1.7 | s2 | 87.34 | 87.31 | 5.3 | 15 |
| | | | | 5130.36 | 0.88 | 3.7 | 1.1 | | 30.01 | | | 16 |
| | | 76.00 | | 5176.11 | 0.52 | 3.0 | 0.7 | | 75.76 | | | 17 |
| | | 36.34 | | 5236.54 | 1.42 | 33.2 | 1.8 | s | 36.18 | > 1 | 85.4 | 18 |
| | | | | 5245.57 | 0.70 | 4.5 | 0.9 | s | 45.21 | 45.27 | 2.7 | 19 |
| | | | | 5247.20 | 1.15 | 6.7 | 1.4 | s | 46.84 | 47.94 | 9.3 | 20 |
| | | | | 5285.62 | 1.07 | 9.3 | 1.3 | s | 85.26 | 84.02 | 12.2 | 21 |
| | | | | 5299.59 | 3.26 | 26.2 | 4.0 | s | 99.23 | 99.41 | 8.4 bl | 22 |
| | | | | 5342.59 | 0.66 | 3.5 | 0.8 | s | 42.23 | 42.31 | 1.1 | 23 |
| | | | | 5358.96 | 1.32 | 5.1 | 1.6 | s | 58.60 | 58.76 | 1.8 | 24 |
| 62.14 | | | | 5361.13 | 7.15 | 118.4 | 8.5 | s; d | 60.77 | 62.81 | 36.9 | 25 |
| 63.60 | 63.60 | 63.78 | | 5364.27 | 1.49 | 29.8 | 1.8 | s | 63.91 | 62.81 | 36.9 | 26 |
| | | 70.36 | | 5370.99 | 1.83 | 8.9 | 2.2 | s | 70.62 | > 1 | 9.9 | 27 |
| 4.56 | 4.50 | 4.52 | | 5404.64 | 1.11 | 52.9 | 1.3 | s | 4.27 | 4.76 | 5.2 | 28 |
| | | | | 5413.52 | 0.50 | 3.8 | 0.6 | | 13.15 | | | 29 |
| | 18.90 | blend | | 5418.96 | 0.83 | 13.1 | 1.0 | | 18.59 | | | 30 |
| | | | | 5420.01 | 12.36 | 166.4 | 14.5 | s | 19.64 | > 1 | 98.0 | 31 |
| 49.63 | | blend | | 5450.52 | 12.30 | 359.5 | 14.4 | s | 50.15 | > 1 | 159.0 | 32 |
| | | | | 5470.85 | 0.44 | 4.3 | 0.5 | | 70.48 | | | 33 |
| blend | 87.67 | blend | | 5487.70 | 3.38 | 235.8 | 3.9 | s | 87.33 | > 1 | 16.1 | 34 |
| 94.14 | 94.10 | 94.10 | | 5494.16 | 0.69 | 31.2 | 0.8 | s | 93.79 | 93.81 | 3.2 | 35 |
| | | | | 5497.65 | 1.25 | < 6.4 | 1.4 | s2 | 97.28 | 95.74 | 4.7 | 36 |
| | | 3.16 | | 5502.94 | 1.18 | 8.4 | 1.4 | s | 2.57 | 3.01 | 5.7 | 37 |
| | | | | 5504.41 | 0.58 | 3.4 | 0.7 | s | 4.04 | 3.01 | 5.7 | 38 |
| 8.35 | 8.35 | 8.03 | | 5508.42 | 2.66 | 158.8 | 3.1 | s; d | 8.04 | > 1 | 104.3 | 39 |
| | 12.64 | 12.66 | | 5512.76 | 0.51 | 8.2 | 0.6 | | 12.38 | | | 40 |
| | | 15.97 | | 5516.02 | 1.30 | 7.3 | 1.5 | | 15.64 | | | 41 |
| 24.89 | | 25.48 | | 5525.21 | 5.48 | < 86.1 | 6.3 | s2 | 24.83 | 26.19 | 11.8 | 42 |
| 35.68 | | blend | | 5535.82 | 3.91 | 139.1 | 4.5 | s | 35.44 | 34.78 | 13.9 | 43 |
| | 41.62 | 41.78 | | 5542.17 | 1.26 | 14.1 | 1.4 | s | 41.79 | 40.94 | 3.6 | 44 |
| 44.97 | 44.96 | 45.02 | | 5545.11 | 0.86 | 28.2 | 1.0 | s | 44.73 | 44.66 | 8.4 | 45 |
| | 46.46 | 46.52 | | 5546.48 | 0.67 | 3.9 | 0.8 | d | 46.10 | | | 46 |
| | | | | 5547.54 | 0.49 | 2.5 | 0.6 | | 47.16 | | | 47 |
| | | 59.93 | | 5560.16 | 1.23 | 6.7 | 1.4 | | 59.78 | | | 48 |
| | | 0.49 | | 5600.60 | 1.63 | 7.2 | 1.8 | | 0.22 | | | 49 |
| 9.96 | 9.73 | 9.96 | | 5610.04 | 1.48 | 27.1 | 1.7 | | 9.66 | | | 50 |
| | | | | 5643.65 | 0.65 | 2.0 | 0.7 | s | 43.27 | 43.70 | 1.6 | 51 |
| | | | | 5652.22 | 1.02 | 5.3 | 1.1 | | 51.83 | | | 52 |
| | | | | 5669.63 | 1.34 | 6.0 | 1.5 | s | 69.24 | 69.44 | 2.1 | 53 |
| | | | | 5674.61 | 0.54 | 2.0 | 0.6 | | 74.22 | | | 54 |
| | 5.20 | 5.10 | 5.43 | 5705.31 | 2.68 | 172.5 | 3.0 | | 4.92 | | | 55 |
| blend | | | | 5705.92 | 14.22 | 279.8 | 15.8 | s; d | 5.53 | > 1 | 71.2 | 56 |
| | | | | 5707.84 | 0.45 | 2.4 | 0.5 | d | 7.45 | | | 57 |
| | | | | 5711.17 | 1.49 | < 36.4 | 1.7 | s2 | 10.78 | 10.73 | 15.2 | 58 |
| 19.43 | 19.30 | 19.68 | 19.40 | 5719.63 | 0.92 | 21.7 | 1.0 | t | 19.24 | | | 59 |
| | | | | 5740.90 | 2.31 | 37.2 | 2.5 | s | 40.51 | 39.72 | 29.6 | 60 |
| | | | | 5744.66 | 1.08 | 20.5 | 1.2 | d | 44.27 | | | 61 |
| | | 46.21 | | 5746.93 | 1.33 | 27.3 | 1.5 | s; d | 46.54 | 47.65 | 9.4 bl | 62 |
| 47.81 | | | | 5748.62 | 2.11 | 27.9 | 2.3 | s; d | 48.23 | 47.65 | 9.4 bl | 63 |
| 62.50 | 62.70 | 62.80 | 62.69 | 5762.70 | 0.61 | 5.8 | 0.7 | | 62.31 | | | 64 |
| 66.25 | 66.16 | 66.17 | 66.15 | 5766.25 | 0.79 | 16.9 | 0.9 | d | 65.86 | | | 65 |
| | | | | 5766.98 | 4.50 | 42.9 | 4.9 | d | 66.59 | | | 66 |
| 72.60 | 72.60 | 72.67 | 72.53 | 5772.66 | 1.23 | 40.4 | 1.4 | d | 72.27 | | | 67 |
| 76.08 | 75.78 | 76.21 | 75.75 | 5775.97 | 0.91 | 25.3 | 1.0 | d | 75.58 | | | 68 |
| 80.59 | 80.37 | 80.55 | 80.50 | 5780.61 | 2.14 | 779.3 | 2.3 | s; nc | 80.22 | 80.13 | 19.5 | 69 |
| 84.90 | 85.05 | 84.86 | 85.11 | 5785.09 | 0.75 | 10.1 | 0.8 | s | 84.70 | > 1 | 8.3 | 70 |
| 89.06 | | 89.04 | 88.90 | 5788.76 | 1.76 | 32.9 | 1.9 | | 88.37 | | | 71 |
| | 93.22 | 93.13 | 93.19 | 5793.17 | 0.87 | 8.2 | 1.0 | | 92.78 | | | 72 |
| 95.23 | 95.16 | | 95.20 | 5795.21 | 1.12 | 12.2 | 1.2 | | 94.82 | | | 73 |
| 97.11 | 96.96 | 97.08 | 96.97 | 5797.20 | 0.91 | 186.4 | 1.0 | nc | 96.81 | | | 74 |
| | | | | 5801.73 | 1.46 | 11.2 | 1.6 | s | 1.33 | 0.43 | 12.9 | 75 |
| blend | 9.24 | 9.53 | 9.22 | 5809.53 | 1.37 | 32.1 | 1.5 | | 9.13 | | | 76 |
| | 11.96 | | 11.57 | 5811.91 | 1.06 | 16.8 | 1.2 | | 11.51 | | | 77 |
| | 15.71 | 15.80 | 15.78 | 5815.72 | 0.85 | 4.9 | 0.9 | s | 15.32 | 15.95 | 3.6 | 78 |
| 18.85 | 18.75 | 18.47 | 18.69 | 5818.81 | 0.65 | 7.7 | 0.7 | s | 18.41 | 19.30 | 14.4 | 79 |
| 28.40 | 28.46 | 28.56 | 28.52 | 5828.63 | 0.88 | 11.5 | 1.0 | s | 28.23 | 27.60 | 2.8 | 80 |
| | 38.00 | 37.92 | 38.08 | 5838.09 | 0.46 | 1.9 | 0.5 | | 37.69 | | | 81 |
| | 40.65 | 40.62 | 40.72 | 5840.69 | 0.42 | 2.0 | 0.5 | | 40.29 | | | 82 |
| 44.19 | | 44.28 | 44.15 | 5843.92 | 4.38 | 100.7 | 4.8 | s | 43.52 | 45.95 | 10.2 bl | 83 |
| | 44.80 | 44.95 | 44.80 | 5844.96 | 0.63 | 11.6 | 0.7 | | 44.56 | | | 84 |
| 49.78 | 49.80 | 49.81 | 49.85 | 5849.88 | 0.93 | 67.8 | 1.0 | | 49.48 | | | 85 |
| | | | 61.89 | 5862.26 | 0.50 | 3.1 | 0.5 | | 61.86 | | | 86 |
| | | | 66.50 | 5866.36 | 0.70 | 5.1 | 0.8 | | 65.96 | | | 87 |
| | 0.40 | 0.58 | 0.56 | 5900.63 | 0.90 | 13.1 | 1.0 | | 0.23 | | | 88 |
| | 10.54 | 10.54 | 10.40 | 5910.61 | 0.73 | 11.4 | 0.8 | | 10.21 | | | 89 |
| | | | 17.51 | 5917.23 | 0.97 | 7.0 | 1.0 | | 16.83 | | | 90 |
| | 23.40 | 23.51 | 23.39 | 5923.62 | 0.82 | 30.4 | 0.9 | | 23.22 | | | 91 |
| | 25.90 | 25.81 | 25.85 | 5926.05 | 0.83 | 21.0 | 0.9 | | 25.65 | | | 92 |
| | 27.68 | | 27.70 | 5927.77 | 0.68 | < 9.1 | 0.7 | s2 | 27.37 | 27.83 | 4.3 | 93 |
| | 47.29 | 47.29 | 47.25 | 5947.31 | 0.78 | 14.9 | 0.8 | | 46.91 | | | 94 |
| | 48.86 | 48.87 | 48.88 | 5948.92 | 0.56 | 5.9 | 0.6 | | 48.52 | | | 95 |
| | | | | 5951.33 | 1.08 | 9.6 | 1.2 | d | 50.92 | | | 96 |
| | | | | 5952.38 | 0.79 | < 12.9 | 0.8 | s2; d | 51.97 | 52.48 | 10.7 bl | 97 |
| | | | | 5953.32 | 0.80 | < 10.9 | 0.9 | s2; d | 52.91 | 52.48 | 10.7 bl | 98 |
| | 58.90 | 58.89 | 58.89 | 5959.00 | -- | p | -- | s | 58.59 | 57.53 | 78.6 bl | 99 |
| | 73.75 | 73.78 | 73.78 | 5973.87 | 0.71 | 5.2 | 0.8 | | 73.46 | | | 100 |
| 82.00 | 82.93 | 82.77 | 82.83 | 5982.81 | 0.85 | 12.1 | 0.9 | | 82.40 | | | 101 |
| | 86.66 | 86.60 | 86.43 | 5986.57 | 0.66 | 6.5 | 0.7 | | 86.16 | | | 102 |
| | 88.08 | 88.04 | 88.03 | 5988.22 | 0.87 | 10.9 | 0.9 | | 87.81 | | | 103 |
| | 95.75 | 95.73 | 95.73 | 5995.93 | 0.83 | 9.3 | 0.9 | s | 95.52 | 96.14 | 6.9 | 104 |
| | 99.63 | | 99.18 | 5999.56 | 0.54 | 7.6 | 0.6 | s; d | 99.15 | 99.61 | 13.9 | 105 |
| | | | 0.78 | 6000.80 | 0.53 | 2.8 | 0.6 | s; d | 0.39 | 99.77 | 3.8 | 106 |
| 4.55 | 5.03 | 4.80 | 4.93 | 6005.04 | 2.08 | 46.5 | 2.2 | | 4.63 | | | 107 |
| 10.58 | 10.65 | 10.48 | 10.80 | 6010.65 | 4.01 | 202.8 | 4.3 | s; d | 10.24 | 12.78 | 3.0 | 108 |
| 19.34 | 19.36 | 19.45 | 19.23 | 6019.39 | 0.93 | 18.6 | 1.0 | s | 18.98 | 18.98 | 2.1 | 109 |
| | | | | 6020.37 | 0.76 | 11.9 | 0.8 | | 19.96 | | | 110 |
| | | | | 6023.45 | 1.29 | 15.0 | 1.4 | s | 23.04 | 24.06 | 16.3 | 111 |
| 27.39 | 27.48 | 27.09 | 27.48 | 6027.70 | 2.08 | 57.8 | 2.2 | s | 27.29 | 26.87 | 4.5 | 112 |
| 37.56 | 37.61 | 37.47 | 37.38 | 6037.54 | 3.88 | 77.3 | 4.1 | s | 37.13 | 37.00 | 7.2 bl | 113 |
| | | | | 6046.76 | 0.99 | 5.3 | 1.1 | | 46.35 | | | 114 |
| | | | | 6051.58 | 0.98 | 5.4 | 1.0 | | 51.17 | | | 115 |
| | | | 57.58 | 6057.61 | 0.68 | 5.1 | 0.7 | | 57.20 | | | 116 |
| | 59.67 | 59.88 | 59.28 | 6059.41 | 0.91 | 10.6 | 1.0 | d | 59.00 | | | 117 |
| 60.41 | | | 60.10 | 6060.38 | 0.69 | 10.1 | 0.7 | d | 59.97 | | | 118 |
| 65.38 | 65.20 | 65.19 | 65.23 | 6065.31 | 0.60 | 13.9 | 0.6 | d | 64.90 | | | 119 |
| | | | | 6065.98 | 2.81 | 25.9 | 3.0 | d | 65.57 | | | 120 |
| | 68.33 | 68.45 | 68.23 | 6068.39 | 1.85 | 16.9 | 2.0 | d | 67.98 | | | 121 |
| | | 71.08 | 70.98 | 6071.22 | 1.16 | 8.8 | 1.2 | | 70.81 | | | 122 |
| | | | | 6081.19 | 0.45 | 2.8 | 0.5 | | 80.78 | | | 123 |
| | 84.75 | 84.94 | 84.78 | 6085.09 | 1.29 | 8.9 | 1.4 | s | 84.68 | 85.87 | 3.5 | 124 |
| | | | | 6087.52 | 0.89 | 8.1 | 0.9 | s | 87.11 | > 1 | 11.1 | 125 |
| 89.80 | 89.78 | 89.79 | 89.78 | 6089.89 | 0.63 | 23.7 | 0.7 | | 89.48 | | | 126 |
| | 2.38 | 2.48 | 2.33 | 6102.43 | 0.64 | 3.5 | 0.7 | | 2.01 | | | 127 |
| 8.21 | 8.05 | 8.14 | 8.13 | 6108.19 | 0.60 | 12.7 | 0.6 | | 7.77 | | | 128 |
| | | | 9.98 | 6109.95 | 0.72 | 7.5 | 0.8 | d | 9.53 | | | 129 |
| | | | | 6110.79 | 0.57 | 2.6 | 0.6 | d | 10.37 | | | 130 |
| 13.22 | 13.20 | 13.20 | 13.18 | 6113.29 | 0.92 | 41.5 | 1.0 | | 12.87 | | | 131 |
| 16.65 | 16.80 | 16.74 | 16.83 | 6116.84 | 0.91 | 13.8 | 1.0 | | 16.42 | | | 132 |
| | 18.68 | 18.66 | 18.63 | 6118.74 | 0.65 | 4.4 | 0.7 | | 18.32 | | | 133 |
| | | | | 6124.43 | 1.08 | 4.2 | 1.1 | | 24.01 | | | 134 |
| | | | | 6128.20 | 2.18 | 11.3 | 2.3 | | 27.78 | | | 135 |
| | | | | 6133.55 | 0.80 | 4.1 | 0.8 | | 33.13 | | | 136 |
| | | | | 6135.98 | 1.41 | 6.1 | 1.5 | | 35.56 | | | 137 |
| 39.77 | 39.94 | 39.94 | 40.03 | 6140.04 | 0.83 | 18.4 | 0.9 | | 39.62 | | | 138 |
| | 41.91 | | | 6142.20 | 0.88 | 6.9 | 0.9 | s | 41.78 | 43.00 | 65.8 | 139 |
| | 45.69 | 45.41 | 45.50 | 6145.74 | 0.47 | 2.4 | 0.5 | | 45.32 | | | 140 |
| | | 51.15 | | 6151.10 | 1.12 | 11.1 | 1.2 | s | 50.68 | 49.20 | 17.4 | 141 |
| | | | | 6159.80 | 1.17 | 11.6 | 1.2 | s | 59.38 | 58.13 | 43.4 | 142 |
| | 61.93 | 61.83 | 61.96 | 6161.98 | 0.72 | 15.9 | 0.8 | | 61.56 | | | 143 |
| | 65.97 | 65.72 | | 6165.62 | 1.90 | 24.5 | 2.0 | s | 65.20 | 63.58 | 19.8 | 144 |
| | 67.84 | | | 6168.08 | 1.06 | < 5.3 | 1.1 | s2 | 67.66 | 67.83 | 2.7 | 145 |
| | | 70.71 | | 6170.81 | 1.96 | 17.1 | 2.1 | | 70.39 | | | 146 |
| | | | 82.60 | 6182.78 | 0.65 | 4.5 | 0.7 | | 82.36 | | | 147 |
| | 85.81 | 85.89 | 85.98 | 6185.86 | 0.57 | 9.2 | 0.6 | | 85.44 | | | 148 |
| | | 87.19 | 87.58 | 6187.47 | 1.12 | 14.4 | 1.2 | | 87.05 | | | 149 |
| 89.53 | | 89.31 | 89.47 | 6189.55 | 0.73 | 7.6 | 0.8 | | 89.13 | | | 150 |
| 94.87 | 94.73 | blend | 94.77 | 6194.79 | 0.57 | 11.4 | 0.6 | | 94.37 | | | 151 |
| 96.19 | 95.96 | 95.99 | blend | 6196.09 | 0.66 | 90.4 | 0.7 | | 95.67 | | | 152 |
| 99.21 | 98.87 | 98.82 | 99.04 | 6199.04 | 0.92 | 10.8 | 1.0 | | 98.62 | | | 153 |
| 3.19 | 3.08 | 3.06 | 3.05 | 6203.14 | 1.42 | 206.2 | 1.5 | d | 2.72 | | | 154 |
| 4.33 | 4.66 | 4.22 | 4.86 | 6205.20 | 2.47 | 151.3 | 2.6 | d | 4.78 | | | 155 |
| | | | | 6210.63 | 1.50 | 6.9 | 1.6 | d | 10.21 | | | 156 |
| 12.19 | 11.67 | 11.74 | 11.73 | 6211.80 | 0.73 | 16.2 | 0.8 | | 11.38 | | | 157 |
| | 12.90 | 12.95 | 12.87 | 6213.00 | 0.64 | 7.8 | 0.7 | | 12.58 | | | 158 |
| 15.71 | 15.79 | 15.80 | | 6215.43 | 1.62 | 10.4 | 1.7 | s | 15.01 | 14.55 | 2.6 | 159 |
| | 20.81 | 20.86 | 20.77 | 6221.02 | 0.88 | 8.1 | 0.9 | s | 20.60 | 19.87 | 3.6 | 160 |
| 23.65 | 23.56 | 23.53 | 23.63 | 6223.66 | 0.51 | 10.3 | 0.5 | s | 23.24 | 22.71 | 4.1 | 161 |
| | | | | 6225.05 | 0.78 | 2.1 | 0.8 | s | 24.63 | 25.31 | 2.9 | 162 |
| | 26.30 | 26.02 | 26.08 | 6226.29 | 0.66 | 5.2 | 0.7 | s | 25.87 | 25.31 | 2.9 | 163 |
| 34.27 | 34.03 | 34.01 | 34.05 | 6234.11 | 0.65 | 18.9 | 0.7 | | 33.69 | | | 164 |
| 36.58 | 36.67 | 36.71 | | 6236.86 | 0.70 | 10.1 | 0.7 | s | 36.44 | 38.34 | 15.4 | 165 |
| | | blend | 45.47 | 6245.09 | 1.36 | 18.8 | 1.4 | | 44.66 | | | 166 |
| | 50.84 | 50.77 | 50.83 | 6250.97 | 1.01 | 11.2 | 1.1 | | 50.54 | | | 167 |
| blend | 69.75 | blend | | 6269.93 | 1.32 | 256.4 | 1.4 | | 69.50 | | | 168 |
| 78.29 | | 78.17 | | 6277.87 | 1.35 | 27.6 | 1.4 | d | 77.44 | | | 169 |
| 80.52 | | 80.48 | | 6280.35 | 0.99 | 22.3 | 1.0 | d | 79.92 | | | 170 |
| 84.31 | 83.85 | blend | 83.80 | 6284.28 | 4.02 | 1884.2 | 4.2 | s; t | 83.85 | 86.85 | 33.4 | 171 |
| 89.60 | 89.74 | 89.70 | 89.86 | 6289.55 | 1.63 | 19.2 | 1.7 | d | 89.12 | | | 172 |
| 8.93 | 9.10 | 8.92 | | 6308.90 | 2.90 | 86.4 | 3.1 | s | 8.47 | 5.42 | 51.5 | 173 |
| 17.58 | 17.06 | 17.75 | | 6316.70 | 2.23 | 45.7 | 2.4 | | 16.27 | | | 174 |
| 25.10 | 24.80 | 24.81 | | 6324.98 | 0.84 | 19.8 | 0.9 | t | 24.55 | | | 175 |
| 30.42 | 29.97 | 30.14 | | 6330.05 | 0.73 | 17.9 | 0.8 | | 29.62 | | | 176 |
| | | | | 6338.02 | 0.50 | 2.8 | 0.5 | | 37.59 | | | 177 |
| | | | 50.70 | 6350.70 | 0.58 | 4.1 | 0.6 | | 50.27 | | | 178 |
| 53.49 | 53.34 | 53.18 | 53.25 | 6353.59 | 1.87 | 51.5 | 2.0 | | 53.16 | | | 179 |
| | | | | 6357.46 | 1.07 | 5.8 | 1.1 | d | 57.03 | | | 180 |
| | | 58.54 | 58.35 | 6358.51 | 0.80 | 9.2 | 0.8 | d | 58.08 | | | 181 |
| 62.35 | 62.30 | 62.23 | 62.50 | 6362.44 | 1.50 | 27.2 | 1.6 | | 62.01 | | | 182 |
| | | | | 6365.07 | 1.23 | 5.7 | 1.3 | d | 64.64 | | | 183 |
| 67.22 | 67.25 | 67.28 | 67.41 | 6367.38 | 0.60 | 17.5 | 0.6 | | 66.95 | | | 184 |
| 76.07 | 75.95 | blend | 76.10 | 6376.21 | 0.94 | 63.7 | 1.0 | | 75.78 | | | 185 |
| 79.27 | 79.29 | 79.22 | 79.46 | 6379.32 | 0.68 | 105.4 | 0.7 | s2 | 78.89 | 79.61 | 5.3 | 186 |
| 97.39 | 97.39 | 96.95 | 96.63 | 6396.88 | 1.03 | 26.4 | 1.1 | s | 96.45 | 97.48 | 24.9 | 187 |
| | 0.30 | 0.37 | 0.25 | 6400.60 | 0.61 | 7.7 | 0.6 | | 0.16 | | | 188 |
| | 10.18 | 10.08 | | 6410.35 | 0.98 | 11.1 | 1.0 | d | 9.91 | | | 189 |
| 13.50 | | | | 6413.17 | 8.28 | 118.2 | 8.8 | s; d | 12.73 | > 1 | 32.9 | 190 |
| 13.90 | 13.93 | blend | 14.18 | 6414.09 | 0.70 | 8.3 | 0.7 | s | 13.65 | 13.52 | 7.6 | 191 |
| | | 18.54 | | 6418.71 | 0.67 | 5.8 | 0.7 | | 18.27 | | | 192 |
| | | | 20.77 | 6420.80 | 0.91 | 4.9 | 1.0 | t | 20.36 | | | 193 |
| 25.72 | 25.70 | 25.61 | 25.90 | 6425.78 | 0.70 | 25.8 | 0.7 | s | 25.34 | 25.79 | 4.6 bl | 194 |
| | | | | 6427.80 | 1.08 | 5.3 | 1.1 | | 27.36 | | | 195 |
| | | | | 6438.28 | 0.65 | 4.7 | 0.7 | d | 37.84 | | | 196 |
| 39.34 | 39.50 | 39.42 | 39.50 | 6439.62 | 0.74 | 26.9 | 0.8 | | 39.18 | | | 197 |
| | | | | 6441.98 | 1.73 | 16.6 | 1.8 | | 41.54 | | | 198 |
| 45.53 | 45.20 | 45.25 | 45.25 | 6445.41 | 0.84 | 58.1 | 0.9 | s | 44.97 | > 1 | 22.1 | 199 |
| 49.13 | 49.14 | 49.30 | 49.28 | 6449.35 | 1.17 | 26.6 | 1.2 | | 48.91 | | | 200 |
| | | | | 6452.22 | 0.45 | 2.0 | 0.5 | | 51.78 | | | 201 |
| | | 56.08 | 55.86 | 6456.02 | 1.01 | 54.5 | 1.1 | s | 55.58 | > 1 | 69.9 | 202 |
| 60.29 | 60.00 | 60.31 | 60.28 | 6460.49 | 0.84 | 10.1 | 0.9 | | 60.05 | | | 203 |
| | 63.61 | 63.63 | 63.41 | 6463.60 | 1.28 | 30.8 | 1.4 | t | 63.16 | | | 204 |
| | 65.48 | | 65.44 | 6465.50 | 0.64 | 5.1 | 0.7 | | 65.06 | | | 205 |
| | 66.74 | 66.95 | 66.65 | 6466.93 | 0.69 | 14.8 | 0.7 | d | 66.49 | | | 206 |
| | 68.70 | 68.45 | 68.53 | 6468.75 | 0.92 | 10.1 | 1.0 | d | 68.31 | | | 207 |
| | 74.27 | | 74.23 | 6474.26 | 1.25 | 12.4 | 1.3 | t | 73.82 | | | 208 |
| 91.88 | 92.02 | 92.17 | 92.15 | 6492.14 | 0.74 | 20.1 | 0.8 | | 91.70 | | | 209 |
| 94.17 | | 92.92 | | 6494.14 | 9.20 | 454.8 | 9.7 | | 93.70 | | | 210 |
| 97.55 | | 97.79 | 97.82 | 6498.01 | 0.79 | 10.4 | 0.8 | | 97.57 | | | 211 |
| | | | 13.38 | 6513.76 | 1.29 | 12.1 | 1.4 | | 13.32 | | | 212 |
| 20.70 | 20.56 | 20.95 | | 6520.75 | 1.02 | 50.6 | 1.1 | | 20.31 | | | 213 |
| | | | 34.14 | 6534.54 | 12.69 | 329.2 | 13.5 | s | 34.10 | 32.87 | 16.6 | 214 |
| 36.44 | | 36.86 | 36.89 | 6536.51 | 1.35 | 20.2 | 1.4 | | 36.07 | | | 215 |
| | 43.20 | | | 6543.29 | 0.83 | 13.5 | 0.9 | | 42.84 | | | 216 |
| | 49.07 | | 48.93 | 6549.09 | 0.71 | 10.2 | 0.8 | | 48.64 | | | 217 |
| | 53.82 | 53.76 | 53.89 | 6553.95 | 0.57 | 17.5 | 0.6 | | 53.50 | | | 218 |
| | 72.84 | | | 6573.04 | 1.15 | 11.6 | 1.2 | | 72.59 | | | 219 |
| | 94.13 | | | 6594.36 | 0.79 | 8.4 | 0.8 | | 93.91 | | | 220 |
| 97.39 | 97.31 | 97.47 | 97.34 | 6597.43 | 0.69 | 26.7 | 0.7 | | 96.98 | | | 221 |
| | | | | 6600.09 | 0.65 | 6.0 | 0.7 | s | 99.64 | 98.91 | 18.3 | 222 |
| | | | | 6607.09 | 0.60 | 4.7 | 0.6 | | 6.64 | | | 223 |
| | | | | 6611.06 | 1.06 | < 18.4 | 1.1 | s2 | 10.61 | > 1 | 9.1 | 224 |
| 13.72 | 13.56 | 13.63 | 13.52 | 6613.70 | 1.08 | 341.6 | 1.2 | | 13.25 | | | 225 |
| | | | | 6621.71 | 0.67 | 5.5 | 0.7 | | 21.26 | | | 226 |
| | | 22.59 | 22.90 | 6622.92 | 0.71 | 17.1 | 0.8 | | 22.47 | | | 227 |
| | | | | 6624.93 | 0.61 | 5.6 | 0.7 | d | 24.48 | | | 228 |
| | | | | 6625.79 | 0.67 | 6.5 | 0.7 | d | 25.34 | | | 229 |
| | 30.80 | | | 6630.82 | 0.56 | 5.3 | 0.6 | d | 30.37 | | | 230 |
| | 31.66 | | | 6631.71 | 0.48 | 4.6 | 0.5 | d | 31.26 | | | 231 |
| 32.93 | 32.85 | 32.86 | 32.65 | 6632.93 | 1.13 | 21.6 | 1.2 | d | 32.48 | | | 232 |
| | | 35.50 | 35.44 | 6635.66 | 0.93 | 9.7 | 1.0 | | 35.21 | | | 233 |
| | | | | 6637.82 | 0.63 | 3.2 | 0.7 | | 37.37 | | | 234 |
| | | 39.34 | | 6639.42 | 0.80 | 5.3 | 0.9 | | 38.97 | | | 235 |
| | | | | 6643.64 | 1.76 | < 32.3 | 1.9 | s1; d | 43.19 | 44.87 | 7.8 bl | 236 |
| | 46.03 | 45.95 | | 6646.05 | 1.16 | < 15.2 | 1.2 | s1; d | 45.60 | 44.87 | 7.8 bl | 237 |
| | 54.58 | | | 6654.63 | 1.12 | 11.5 | 1.2 | | 54.18 | | | 238 |
| | | | | 6657.47 | 1.08 | 9.4 | 1.2 | | 57.02 | | | 239 |
| | | | | 6658.77 | 0.67 | 6.8 | 0.7 | | 58.32 | | | 240 |
| 60.64 | 60.64 | 60.62 | 60.70 | 6660.73 | 0.67 | 59.7 | 0.7 | s | 60.28 | 60.53 | 29.4 | 241 |
| | | 62.25 | | 6662.29 | 0.65 | 6.1 | 0.7 | s | 61.84 | 60.53 | 29.4 | 242 |
| | | | | 6664.05 | 1.07 | 7.6 | 1.2 | s | 63.60 | 64.95 | 13.9 | 243 |
| | | | | 6669.51 | 1.17 | 9.2 | 1.3 | | 69.06 | | | 244 |
| | | 84.83 | | 6684.91 | 1.64 | 14.0 | 1.8 | | 84.46 | | | 245 |
| | 89.30 | 89.38 | | 6689.38 | 0.99 | 21.3 | 1.1 | | 88.92 | | | 246 |
| | 93.35 | | | 6693.63 | 0.73 | 6.4 | 0.8 | d | 93.17 | | | 247 |
| 94.46 | 94.48 | 94.40 | 94.50 | 6694.60 | 0.65 | 7.2 | 0.7 | d | 94.14 | | | 248 |
| | | | 97.01 | 6697.06 | 0.73 | 5.9 | 0.8 | | 96.60 | | | 249 |
| 99.37 | 99.24 | 99.28 | 99.24 | 6699.36 | 0.82 | 43.3 | 0.9 | s | 98.90 | 99.28 | 10.1 | 250 |
| 1.98 | 1.98 | 1.87 | 1.95 | 6702.15 | 0.74 | 17.1 | 0.8 | | 1.69 | | | 251 |
| | | | | 6706.61 | 0.96 | < 6.5 | 1.0 | s | 6.15 | 4.83 | 9.5 bl | 252 |
| 9.24 | 9.39 | 9.65 | 9.44 | 6709.54 | 0.96 | 18.7 | 1.0 | s | 9.08 | 9.09 | 2.4 | 253 |
| | | | | 6713.81 | 1.13 | 10.5 | 1.2 | | 13.35 | | | 254 |
| | | | | 6718.38 | 0.62 | 5.3 | 0.7 | s; d | 17.92 | 16.98 | 37.0 | 255 |
| | | 19.58 | | 6719.25 | 0.71 | 7.7 | 0.8 | d | 18.79 | | | 256 |
| | | | | 6727.68 | 0.71 | 7.6 | 0.8 | d | 27.22 | | | 257 |
| | | | | 6728.55 | 0.70 | 7.7 | 0.8 | d | 28.09 | | | 258 |
| | 29.28 | 29.20 | 29.28 | 6729.30 | 0.67 | 8.4 | 0.7 | d | 28.84 | | | 259 |
| | | | | 6731.21 | 0.92 | 3.9 | 1.0 | | 30.75 | | | 260 |
| | | 33.35 | | 6733.38 | 1.24 | 10.7 | 1.3 | s; d | 32.92 | 33.11 | 2.9 | 261 |
| | | | | 6735.23 | 3.64 | 29.0 | 3.9 | d | 34.77 | | | 262 |
| | 37.13 | | 37.14 | 6737.40 | 0.81 | 10.9 | 0.9 | d | 36.94 | | | 263 |
| 40.96 | 40.99 | | 40.96 | 6741.13 | 1.19 | 20.7 | 1.3 | d | 40.67 | | | 264 |
| | | | | 6742.58 | 0.89 | 11.3 | 1.0 | d | 42.12 | | | 265 |
| | | | 44.06 | 6743.78 | 1.48 | 17.3 | 1.6 | d | 43.32 | | | 266 |
| | | | 47.80 | 6747.91 | 1.19 | 12.3 | 1.3 | | 47.45 | | | 267 |
| | | | | 6752.36 | 1.13 | 13.4 | 1.2 | s | 51.90 | 50.11 | 8.5 | 268 |
| | | | | 6755.99 | 1.06 | 7.9 | 1.2 | d | 55.53 | | | 269 |
| | | | | 6757.18 | 0.90 | 6.5 | 1.0 | d | 56.72 | | | 270 |
| | | | | 6759.28 | 1.09 | 5.0 | 1.2 | | 58.82 | | | 271 |
| | | | | 6762.29 | 1.13 | 7.7 | 1.2 | | 61.83 | | | 272 |
| | | 65.39 | | 6765.38 | 0.88 | 5.2 | 1.0 | | 64.92 | | | 273 |
| | 67.74 | | 67.58 | 6767.69 | 0.71 | 5.1 | 0.8 | | 67.23 | | | 274 |
| 70.05 | 70.05 | 70.23 | 70.12 | 6770.23 | 0.82 | 22.1 | 0.9 | | 69.77 | | | 275 |
| | | | | 6773.36 | 0.66 | 4.4 | 0.7 | d | 72.90 | | | 276 |
| | | | | 6775.44 | 0.72 | 4.0 | 0.8 | d | 74.98 | | | 277 |
| | 78.99 | | 79.04 | 6779.05 | 0.64 | 5.2 | 0.7 | | 78.59 | | | 278 |
| | | | | 6786.55 | 0.86 | 6.2 | 0.9 | | 86.09 | | | 279 |
| 88.89 | 88.66 | | 88.66 | 6788.84 | 1.02 | 14.3 | 1.1 | | 88.38 | | | 280 |
| 92.45 | 92.52 | 92.54 | 92.42 | 6792.55 | 0.80 | 16.2 | 0.9 | | 92.09 | | | 281 |
| 95.37 | 95.24 | 95.18 | 95.20 | 6795.29 | 0.62 | 10.6 | 0.7 | d | 94.83 | | | 282 |
| | | | | 6796.13 | 1.19 | 7.1 | 1.3 | d | 95.67 | | | 283 |
| 1.41 | 1.37 | 1.39 | 1.41 | 6801.54 | 0.80 | 17.4 | 0.9 | | 1.08 | | | 284 |
| 3.28 | 3.29 | | 3.26 | 6803.37 | 0.60 | 9.1 | 0.7 | d | 2.91 | | | 285 |
| | | 4.82 | | 6804.68 | 1.38 | 6.8 | 1.5 | d | 4.22 | | | 286 |
| | | | | 6809.53 | 0.81 | 5.8 | 0.9 | | 9.07 | | | 287 |
| 11.44 | | 11.30 | 11.27 | 6811.31 | 0.93 | 27.6 | 1.0 | d | 10.85 | | | 288 |
| 12.82 | | 12.39 | 12.80 | 6812.76 | 1.41 | 25.1 | 1.5 | d | 12.30 | | | 289 |
| | | | | 6818.31 | 1.49 | 9.5 | 1.6 | | 17.85 | | | 290 |
| | 21.56 | | | 6821.69 | 1.32 | 13.8 | 1.4 | s | 21.23 | 22.40 | 6.9 | 291 |
| | 23.30 | | 23.38 | 6823.60 | 0.56 | 6.5 | 0.6 | s | 23.14 | 22.40 | 6.9 | 292 |
| | | | | 6825.83 | 0.90 | 5.5 | 1.0 | d | 25.37 | | | 293 |
| 27.28 | 27.30 | | 27.22 | 6827.44 | 1.04 | 29.7 | 1.1 | | 26.98 | | | 294 |
| | | | | 6831.21 | 0.60 | 1.8 | 0.7 | | 30.75 | | | 295 |
| | 34.50 | | 34.53 | 6834.60 | 0.86 | 9.2 | 0.9 | | 34.14 | | | 296 |
| | | | | 6836.32 | 0.71 | 4.8 | 0.8 | | 35.85 | | | 297 |
| | 37.70 | | 37.70 | 6837.75 | 0.72 | 8.4 | 0.8 | | 37.28 | | | 298 |
| | | | | 6839.56 | 0.47 | 2.2 | 0.5 | | 39.09 | | | 299 |
| 41.59 | 41.49 | | 41.54 | 6841.65 | 0.59 | 7.6 | 0.6 | | 41.18 | | | 300 |
| 43.44 | 43.60 | | 43.69 | 6843.64 | 1.24 | 46.4 | 1.4 | | 43.17 | | | 301 |
| | 45.30 | | 45.26 | 6845.40 | 0.67 | 6.4 | 0.7 | | 44.93 | | | 302 |
| | 46.60 | | 46.53 | 6846.62 | 0.51 | 4.4 | 0.6 | | 46.15 | | | 303 |
| | 47.76 | | | 6847.78 | 0.76 | 6.5 | 0.8 | | 47.31 | | | 304 |
| 52.90 | 52.67 | | 52.91 | 6852.54 | 0.80 | 15.7 | 0.9 | | 52.07 | | | 305 |
| 60.02 | 60.02 | | 60.13 | 6860.17 | 0.99 | 37.4 | 1.1 | d | 59.70 | | | 306 |
| | 62.53 | | 62.47 | 6862.61 | 0.65 | 11.6 | 0.7 | d | 62.14 | | | 307 |
| | 64.65 | | | 6864.67 | 0.45 | 2.0 | 0.5 | | 64.20 | | | 308 |
| 86.92 | 86.56 | | | 6886.97 | 1.32 | 77.7 | 1.4 | t | 86.50 | | | 309 |
| 19.25 | 19.44 | | | 6919.28 | 1.05 | 64.9 | 1.1 | | 18.81 | | | 310 |
| | | | | 6926.58 | 0.88 | 18.8 | 1.0 | | 26.11 | | | 311 |
| 44.53 | 44.56 | | | 6944.80 | 0.86 | 30.8 | 0.9 | | 44.33 | | | 312 |
| | 50.55 | | | 6950.71 | 0.74 | 11.5 | 0.8 | | 50.24 | | | 313 |
| | | | | 6951.81 | 0.86 | 8.2 | 0.9 | | 51.34 | | | 314 |
| | 73.55 | | | 6973.72 | 0.76 | 18.9 | 0.8 | | 73.25 | | | 315 |
| 78.54 | 78.28 | | | 6978.55 | 0.80 | 16.1 | 0.9 | | 78.08 | | | 316 |
| 93.18 | 93.18 | | | 6993.24 | 0.94 | 173.1 | 1.0 | | 92.76 | | | 317 |
| | | | | 6996.88 | 0.67 | 5.9 | 0.7 | | 96.40 | | | 318 |
| 98.71 | 98.76 | | | 6998.76 | 0.72 | 13.7 | 0.8 | t | 98.28 | | | 319 |
| | 2.19 | | | 7002.45 | 0.91 | 17.0 | 1.0 | s | 1.97 | 2.13 | 24.3 | 320 |
| | | | | 7004.51 | 0.85 | 15.6 | 0.9 | | 4.03 | | | 321 |
| | | | | 7025.48 | 1.15 | 11.9 | 1.3 | | 25.00 | | | 322 |
| | 30.35 | | | 7030.31 | 0.95 | 20.5 | 1.0 | d | 29.83 | | | 323 |
| | 31.56 | | | 7031.67 | 0.88 | 31.7 | 1.0 | s; d | 31.19 | 32.38 | 41.4 | 324 |
| 45.65 | 45.87 | | | 7045.92 | 0.80 | 16.7 | 0.9 | | 45.44 | | | 325 |
| 60.81 | 61.00 | | | 7061.19 | 0.71 | 29.7 | 0.8 | | 60.71 | | | 326 |
| 62.70 | 62.65 | | | 7062.76 | 0.69 | 39.2 | 0.8 | | 62.28 | | | 327 |
| 69.65 | 69.48 | | | 7069.68 | 1.02 | 43.4 | 1.1 | | 69.20 | | | 328 |
| 78.02 | 78.11 | | | 7078.07 | 0.73 | 14.5 | 0.8 | | 77.59 | | | 329 |
| | | | | 7083.80 | 0.63 | 6.3 | 0.7 | d | 83.32 | | | 330 |
| 85.10 | 84.94 | | | 7085.48 | 1.53 | 21.8 | 1.7 | d | 85.00 | | | 331 |
| | | | | 7086.97 | 0.60 | 4.6 | 0.7 | d | 86.49 | | | 332 |
| 1.14 | | | | 7101.14 | 1.03 | 11.4 | 1.1 | t | 0.66 | | | 333 |
| | | | | 7105.29 | 0.78 | 6.1 | 0.9 | d | 4.81 | | | 334 |
| 5.93 | 5.81 | | | 7106.31 | 3.99 | 63.4 | 4.4 | | 5.83 | | | 335 |
| | | | | 7116.48 | 0.71 | 8.3 | 0.8 | t | 16.00 | | | 336 |

| | | | | | | | | | | |
|---|---|---|---|---|---|---|---|---|---|---|
| 19.94 | | 7120.01 | 1.23 | 39.9 | 1.4 | | 19.53 | | | 337 |
| | | 7124.68 | 1.32 | 8.5 | 1.5 | | 24.19 | | | 338 |
| | | 7136.89 | 1.91 | 30.6 | 2.1 | s; d | 36.40 | 35.43 | 5.3 | 339 |
| blend | | 7138.69 | 0.82 | 7.3 | 0.9 | t | 38.20 | | | 340 |
| | | 7140.12 | 3.78 | 35.1 | 4.2 | s | 39.63 | 39.75 | 3.7 | 341 |
| | | 7143.00 | 0.61 | 4.8 | 0.7 | t | 42.51 | | | 342 |
| | | 7152.25 | 2.26 | 15.0 | 2.5 | d | 51.76 | | | 343 |
| 53.78 | 53.94 | 7154.07 | 1.21 | 12.2 | 1.3 | d; t | 53.58 | | | 344 |
| | 59.51 | 7159.47 | 0.63 | 13.2 | 0.7 | d; t | 58.98 | | | 345 |
| | | 7160.53 | 1.83 | 27.4 | 2.0 | d | 60.04 | | | 346 |
| | 62.96 | 7163.16 | 0.84 | 15.9 | 0.9 | | 62.67 | | | 347 |
| | | 7166.07 | 0.57 | 4.2 | 0.6 | t | 65.58 | | | 348 |
| | | 7203.62 | 0.56 | 20.2 | 0.6 | | 3.13 | | | 349 |
| 24.18 | 24.00 | 7224.18 | 1.24 | 358.8 | 1.4 | | 23.69 | | | 350 |
| 49.26 | 49.28 | 7249.09 | 1.61 | 26.3 | 1.8 | | 48.60 | | | 351 |
| 57.35 | 57.49 | 7257.44 | 1.36 | 22.5 | 1.5 | | 56.95 | | | 352 |
| | | 7268.53 | 1.29 | 19.2 | 1.4 | | 68.04 | | | 353 |
| 76.70 | | 7276.65 | 0.68 | 41.2 | 0.8 | t | 76.16 | | | 354 |
| | | 7322.11 | 0.58 | 12.5 | 0.6 | t | 21.61 | | | 355 |
| 30.17 | | 7329.84 | 1.94 | 27.6 | 2.2 | | 29.34 | | | 356 |
| 34.33 | | 7334.66 | 1.08 | 60.6 | 1.2 | | 34.16 | | | 357 |
| | | 7338.14 | 0.94 | 7.2 | 1.0 | | 37.64 | | | 358 |
| | | 7339.94 | 1.12 | 6.5 | 1.2 | t | 39.44 | | | 359 |
| | | 7342.73 | 1.71 | 10.7 | 1.9 | t | 42.23 | | | 360 |
| | | 7347.50 | 1.36 | 12.1 | 1.5 | | 47.00 | | | 361 |
| 49.81 | 49.79 | 7349.86 | 0.81 | 16.5 | 0.9 | | 49.36 | | | 362 |
| 54.92 | 54.60 | 7354.94 | 0.96 | 14.5 | 1.1 | | 54.44 | | | 363 |
| blend | 57.60 | 7357.65 | 0.58 | 53.4 | 0.6 | | 57.15 | | | 364 |
| 60.45 | 60.49 | 7360.62 | 0.68 | 12.2 | 0.8 | | 60.12 | | | 365 |
| | | 7363.90 | 0.74 | 5.8 | 0.8 | t | 63.40 | | | 366 |
| 66.61 | | 7366.16 | 0.93 | 25.3 | 1.0 | d | 65.66 | | | 367 |
| | 67.12 | 7367.21 | 0.75 | 48.6 | 0.8 | d | 66.71 | | | 368 |
| 75.90 | 75.90 | 7375.93 | 0.81 | 14.8 | 0.9 | | 75.43 | | | 369 |
| | | 7382.62 | 0.80 | 8.6 | 0.9 | | 82.12 | | | 370 |
| 85.92 | 85.83 | 7385.98 | 0.64 | 15.2 | 0.7 | | 85.48 | | | 371 |
| 1.71 | | 7402.36 | 2.32 | 85.1 | 2.6 | | 1.86 | | | 372 |
| 6.35 | 6.30 | 7406.48 | 2.04 | 32.8 | 2.3 | | 5.98 | | | 373 |
| | 19.07 | 7419.13 | 0.84 | 14.7 | 0.9 | | 18.63 | | | 374 |
| | | 7451.46 | 1.00 | 20.5 | 1.1 | s | 50.95 | 49.04 | 6.1 | 375 |
| | 58.15 | 7458.30 | 1.88 | 10.6 | 2.1 | | 57.79 | | | 376 |
| | 70.35 | 7470.39 | 0.60 | 9.3 | 0.7 | s1 | 69.88 | 68.26 | 74.7 | 377 |
| | 72.65 | 7472.63 | 0.48 | 4.1 | 0.5 | | 72.12 | | | 378 |
| | | 7476.77 | 1.11 | 7.3 | 1.2 | | 76.26 | | | 379 |
| | | 7478.47 | 0.73 | 5.2 | 0.8 | | 77.96 | | | 380 |
| | 83.02 | 7483.32 | 2.23 | 25.9 | 2.5 | | 82.81 | | | 381 |
| | | 7493.23 | 1.09 | 12.8 | 1.2 | s(e) | 92.72 | P Cyg | -- | 382 |
| | | 7520.53 | 0.98 | 7.0 | 1.1 | | 20.02 | | | 383 |
| | | 7532.73 | 1.12 | 9.5 | 1.3 | | 32.22 | | | 384 |
| | | 7544.11 | 2.58 | 19.7 | 2.9 | s | 43.60 | 43.73 | 7.1 | 385 |
| | | 7553.77 | 1.34 | 5.4 | 1.5 | | 53.26 | | | 386 |
| 58.50 | | 7558.44 | 1.86 | 21.3 | 2.1 | d | 57.93 | | | 387 |
| | 59.35 | 7559.48 | 0.73 | 15.4 | 0.8 | d | 58.97 | | | 388 |
| 62.24 | | 7562.29 | 1.72 | 100.5 | 1.9 | | 61.78 | | | 389 |
| | | 7564.45 | 0.95 | 16.5 | 1.1 | d | 63.94 | | | 390 |
| | | 7568.22 | 0.87 | 11.7 | 1.0 | d | 67.71 | | | 391 |
| 69.70 | | 7569.83 | 5.61 | 41.3 | 6.3 | d | 69.32 | | | 392 |
| | | 7570.29 | 0.75 | 5.8 | 0.8 | d | 69.78 | | | 393 |
| | 71.66 | 7571.91 | 0.91 | 11.3 | 1.0 | d | 71.40 | | | 394 |
| 79.17 | | 7579.59 | 1.14 | 24.6 | 1.3 | d | 79.08 | | | 395 |
| 81.24 | 81.30 | 7581.41 | 1.50 | 52.7 | 1.7 | d | 80.90 | | | 396 |
| 85.63 | | 7585.66 | 1.40 | 13.8 | 1.6 | | 85.15 | | | 397 |
| 95.92 | 96.00 | 7696.08 | 0.86 | 17.8 | 1.0 | | 95.56 | | | 398 |
| 5.90 | | 7705.90 | 2.32 | 23.1 | 2.6 | | 5.38 | | | 399 |
| | 7.96 | 7708.25 | 0.72 | 11.1 | 0.8 | | 7.73 | | | 400 |
| 21.88 | 21.85 | 7722.07 | 0.74 | 32.8 | 0.8 | | 21.54 | | | 401 |
| | | 7823.17 | 1.92 | 13.6 | 2.2 | | 22.64 | | | 402 |
| | | 7827.74 | 1.35 | 16.2 | 1.5 | | 27.21 | | | 403 |
| 32.72 | 32.81 | 7833.01 | 0.82 | 39.6 | 0.9 | | 32.48 | | | 404 |
| | | 7841.05 | 1.25 | 12.4 | 1.4 | | 40.52 | | | 405 |
| 62.34 | 62.39 | 7862.56 | 0.82 | 15.1 | 0.9 | | 62.03 | | | 406 |
| 35.33 | | 7935.31 | 1.34 | 16.1 | 1.5 | | 34.77 | | | 407 |
| | | 7950.77 | 2.02 | 31.8 | 2.3 | s | 50.23 | 51.18 | 17.7 bl | 408 |
| | | 7968.09 | 1.86 | 16.4 | 2.1 | s | 67.55 | 67.38 | 9.2 | 409 |
| | | 7971.92 | 2.12 | 41.1 | 2.4 | t | 71.38 | | | 410 |
| | | 8001.23 | 1.18 | 12.1 | 1.3 | t | 0.69 | | | 411 |
| 26.21 | 26.27 | 8026.37 | 0.75 | 44.0 | 0.9 | | 25.83 | | | 412 |
| 38.48 | | 8038.58 | 5.37 | 204.7 | 6.1 | t | 38.04 | | | 413 |
| | | 8085.94 | 2.36 | 33.4 | 2.7 | | 85.39 | | | 414 |
| | | 8154.23 | 3.30 | 92.1 | 3.8 | | 53.68 | | | 415 |
| blend | 20.79 | 8620.99 | 3.57 | 355.8 | 4.2 | | 20.40 | | | 416 |
| | | 8763.83 | 0.99 | 21.2 | 1.0 | s | 63.23 | 63.77 | 5.9 | 417 |
| | | 8772.77 | 2.36 | 24.4 | 2.8 | | 72.17 | | | 418 |

Table 3

71 Possible Diffuse Interstellar Bands in the Spectrum of HD 183143

| JD94 $\lambda_c$(A) | GM00 $\lambda_c$(A) | TC00 $\lambda_c$(A) | WS00 $\lambda_c$(A) | $\lambda_c$(A) | FWHM (A) | $W_\lambda$ (mA) | $\Delta W_\lambda$ (mA) | note | DIB $\lambda_c$*(A) | β Ori $\lambda_c$*(A) | β Ori $W_\lambda$ (mA) |
|---|---|---|---|---|---|---|---|---|---|---|---|
| | | | | 3983.60 | 5.31 | 91.1 | 17.8 | s | 83.33 | > 1 | 17.9 |
| | | | | 4675.02 | 1.08 | < 4.9 | 1.7 | s | 74.70 | 74.83 | 2.5 |
| | | | | 4678.50 | 0.74 | 4.8 | 1.2 | | 78.18 | | |
| | | | | 4683.04 | 0.54 | 3.8 | 0.9 | | 82.72 | | |
| | | | | 4705.55 | -- | < 9.7 | -- | s | 5.23 | 5.41 | 3.9 |
| | | | | 4757.89 | 14.45 | 176.4 | 22.0 | s; d | 57.57 | 55.20 | 3.8 |
| | | | | 4810.34 | 1.18 | 7.7 | 1.7 | s | 10.01 | 10.11 | 3.5 |
| 80.35 | | | | 4880.12 | 1.32 | < 17.5 | 1.9 | s; d | 79.79 | 79.83 | 8.5 |
| | | | | 4898.47 | 1.72 | 7.3 | 2.4 | | 98.14 | | |
| | | | | 4933.30 | 1.25 | 4.4 | 1.7 | | 32.96 | | |
| | | | | 5063.67 | 1.20 | 5.6 | 1.6 | s | 63.33 | 63.52 | 7.1 |
| | | | | 5117.46 | 1.35 | < 5.8 | 1.7 | s | 17.11 | 16.85 | 2.9 |
| | | | | 5134.39 | 1.28 | 5.5 | 1.6 | s | 34.04 | 33.02 | 11.7 |
| | | | | 5340.76 | 0.88 | 4.4 | 1.1 | s | 40.40 | 39.61 | 18.7 |
| | | | | 5449.93 | 1.15 | 6.3 | 1.3 | | 49.56 | | |
| | | | | 5462.86 | 0.62 | 4.2 | 0.7 | | 62.49 | | |
| | | | | 5480.83 | 0.47 | 2.1 | 0.5 | | 80.46 | | |
| | | 90.52 | | 5490.44 | 0.52 | 2.3 | 0.6 | d | 90.07 | | |
| | | | | 5590.27 | 0.69 | 2.1 | 0.8 | | 89.89 | | |
| | | 34.73 | | 5635.30 | 1.45 | 5.2 | 1.6 | | 34.92 | | |
| | | | | 5637.38 | 1.23 | 4.5 | 1.4 | | 37.00 | | |
| | | | | 5716.25 | 0.60 | 2.6 | 0.7 | | 15.86 | | |
| | 21.23 | | 21.22 | 5821.26 | 0.45 | 1.1 | 0.5 | | 20.86 | | |
| | 54.50 | | 54.54 | 5854.54 | 0.45 | 2.3 | 0.5 | | 54.14 | | |
| | 55.63 | | 55.72 | 5855.62 | 0.68 | 2.4 | 0.7 | | 55.22 | | |
| | | | | 5954.33 | 0.60 | 5.9 | 0.6 | d | 53.92 | | |
| | | | 65.28 | 5965.35 | 0.68 | 3.5 | 0.7 | s | 64.94 | 65.58 | 1.2 |
| | 75.74 | 75.66 | 75.58 | 5975.74 | 0.40 | 3.4 | 0.4 | s; t | 75.33 | 75.47 | 4.0 |
| | | | | 6035.81 | 0.54 | 1.4 | 0.6 | | 35.40 | | |
| | | | | 6079.11 | 3.51 | 14.9 | 3.7 | d | 78.70 | | |
| | | | 93.18 | 6093.35 | 0.81 | 2.1 | 0.9 | | 92.93 | | |
| | | | | 6252.36 | 0.73 | 2.9 | 0.8 | | 51.93 | | |
| | | | | 6259.84 | 0.84 | 3.6 | 0.9 | s | 59.41 | 59.26 | 5.4bl |
| | | | | 6339.87 | 1.00 | 4.6 | 1.1 | | 39.44 | | |
| | | | | 6355.51 | 0.69 | 2.7 | 0.7 | d | 55.08 | | |
| | | | | 6369.58 | -- | p | -- | s | 69.15 | 71.34 | 484.2 |
| | | | | 6406.39 | 0.57 | 2.2 | 0.6 | | 5.95 | | |
| | 76.94 | 76.81 | | 6477.05 | 0.67 | 3.0 | 0.7 | | 76.61 | | |
| | | | | 6487.22 | 1.17 | 6.1 | 1.2 | s | 86.78 | 87.14 | 4.7 |
| 91.40 | | 91.03 | | 6590.58 | 4.41 | 49.5 | 4.7 | | 90.13 | | |
| | | | | 6618.38 | 0.84 | 4.6 | 0.9 | | 17.93 | | |
| | | | | 6651.90 | 0.69 | 3.9 | 0.7 | | 51.45 | | |
| | 65.15 | | 65.15 | 6665.25 | 0.57 | 4.6 | 0.6 | s | 64.80 | 64.95 | 13.9 |
| | 72.15 | | 72.39 | 6672.30 | 0.96 | < 52.6 | 1.0 | s | 71.85 | 71.78 | 36.7 |
| | | 86.46 | | 6686.75 | 0.60 | 2.3 | 0.6 | | 86.30 | | |
| | | | | 6724.14 | 0.67 | 2.6 | 0.7 | | 23.68 | | |
| | | | | 6774.35 | 0.75 | 4.3 | 0.8 | d | 73.89 | | |
| | | | | 6777.38 | 1.20 | 7.9 | 1.3 | | 76.92 | | |
| | | | | 6781.00 | 0.86 | 3.6 | 0.9 | s | 80.54 | 79.99 | 9.9 |
| | | | | 6783.44 | 0.54 | 2.0 | 0.6 | s | 82.98 | 83.95 | 8.5 |
| | | | | 6791.49 | 0.61 | 2.8 | 0.7 | d | 91.03 | | |
| | | | | 6833.14 | 0.83 | 2.2 | 0.9 | | 32.68 | | |
| | 49.56 | | 49.66 | 6849.95 | 1.12 | 3.5 | 1.2 | | 49.48 | | |
| | | | | 6966.06 | 0.64 | 6.6 | 0.7 | t | 65.59 | | |
| | 71.51 | | | 6971.77 | 1.26 | 8.0 | 1.4 | t | 71.30 | | |
| | 82.46 | | | 6982.63 | 1.07 | 10.6 | 1.2 | s; t | 82.16 | 81.50 | 9.9 |
| | | | | 7172.59 | 0.88 | 13.6 | 1.0 | t | 72.10 | | |
| | | | | 7198.64 | 0.48 | 9.7 | 0.5 | t | 98.15 | | |
| 64.98 | | | | 7264.69 | 0.60 | 11.8 | 0.7 | t | 64.20 | | |
| | | | | 7301.15 | 0.75 | 10.4 | 0.8 | t | 0.65 | | |
| | | | | 7359.20 | 0.60 | 4.7 | 0.7 | d; t | 58.70 | | |
| | | | | 7473.91 | 0.51 | 2.0 | 0.6 | | 73.40 | | |
| | | | | 7525.95 | 0.57 | 4.2 | 0.6 | s | 25.44 | 25.14 | 6.7 |
| | | | | 7706.87 | 0.64 | 6.0 | 0.7 | | 6.35 | | |
| 21.03 | 20.26 | | | 7720.45 | 0.68 | 5.9 | 0.8 | | 19.92 | | |
| | | | | 7829.62 | 1.08 | 5.4 | 1.2 | d | 29.09 | | |
| 4.92 | | | | 7904.44 | -- | p | -- | t | 3.90 | | |
| 15.13 | 15.36 | | | 7915.53 | 0.84 | 5.3 | 1.0 | s; t | 14.99 | 15.23 | 7.3 |
| | | | | 8043.73 | 1.66 | 19.1 | 1.9 | d; t | 43.18 | | |
| | | | | 8046.38 | 1.26 | 11.5 | 1.4 | | 45.83 | | |
| | | | | 8583.86 | 4.63 | 107.9 | 5.4 | s | 83.28 | 82.56 | 70.8bl |

Table 4

Some Statistical Properties of the DIBs toward HD 183143

| DIB # | $\lambda_{min}$(Å) | $\lambda_{max}$(Å) | $\Delta\lambda$(Å) | Sp[a](Å) | New[b] | Prev[c] | New Fr[d] |
|---|---|---|---|---|---|---|---|
| 1-50 | 3900.0 | 5620.0 | 1720 | 34.4 | 20 | 30 | 0.40 |
| 51-100 | 5620.0 | 5980.0 | 360 | 7.2 | 14 | 36 | 0.28 |
| 101-150 | 5980.0 | 6198.0 | 210 | 4.2 | 13 | 37 | 0.26 |
| 151-200 | 6190.0 | 6450.0 | 260 | 5.2 | 8 | 42 | 0.16 |
| 201-250 | 6450.0 | 6700.0 | 250 | 5.0 | 13 | 37 | 0.26 |
| 251-300 | 6700.0 | 6842.0 | 142 | 2.8 | 23 | 27 | 0.46 |
| 301-350 | 6842.0 | 7230.0 | 388 | 7.8 | 17 | 33 | 0.34 |
| 350-400 | 7230.0 | 7715.0 | 485 | 9.7 | 6 | 24 | 0.20 |
| 401-414 | 7715.0 | 8100.0 | 385 | 7.7 | 8 | 6 | 0.57 |
| 1–414 | 3900.0 | 8100.0 | 4200 | 10.1 | 135 | 279 | 0.33 |

a The average spacing between DIBs in each wavelength range.
b The number of “new” DIBs in each wavelength range.
c The number of previously known DIBs in each wavelength range.
d The fraction of “new” DIBs in each wavelength range

Table 5

The Narrowest DIBs toward HD 183143

| $\lambda$ (Å) | FWHM (Å) | $W_\lambda$ (mÅ) | FWHM (km $s^{-1}$) |
|---|---|---|---|
| 5413.52 | 0.50 | 3.8 | 27.7 |
| 5470.82 | 0.44 | 4.3 | 24.1 |
| 5547.54 | 0.49 | 2.5 | 26.5 |
| 5707.84 | 0.45 | 2.4 | 23.6 |
| 5838.09 | 0.46 | 1.9 | 23.6 |
| 5840.69 | 0.42 | 2.0 | 21.6 |
| 5862.26 | 0.50 | 3.1 | 25.6 |
| 6081.19 | 0.45 | 2.8 | 22.2 |
| 6145.74 | 0.47 | 2.4 | 22.9 |
| 6338.02 | 0.50 | 2.8 | 23.7 |
| 6452.22 | 0.45 | 2.0 | 20.9 |
| 6631.71 | 0.48 | 4.6 | 21.7 |
| 6839.56 | 0.47 | 2.2 | 20.6 |
| 6864.67 | 0.45 | 2.0 | 19.7 |
| 7472.63 | 0.48 | 4.1 | 19.3 |

Table 6

Atomic and Diatomic Interstellar Lines

| Absorber | Band | Line | $\lambda$(Å) | $W_\lambda$ (mÅ) |
|---|---|---|---|---|
| Fe I | | | 3859.91 | 4.1 |
| CN | B-X (0-0) | R(1) | 3874.00 | 2.7 |
| CN | B-X (0-0) | R(0) | 3874.61 | 7.1 |
| CN | B-X (0-0) | P(1) | 3875.77 | 1.6 |
| CH | B-X (0-0) | $R_2$(1/2) | 3878.77 | 1.3 |
| CH | B-X (0-0) | $Q_2$(1/2) | 3886.41 | 11.4 |
| CH | B-X (0-0) | $P_2$(1/2) | 3890.21 | 1.8 |
| Ca II | | | 3933.66 | 338. |
| $CH^+$ | A-X (1-0) | R(0) | 3957.71 | 30.6 |
| Ca II | | | 3968.47 | 256. |
| Ca I | | | 4226.73 | 10.7 |
| $CH^+$ | A-X (0-0) | R(0) | 4232.55 | 47.7 |
| CH | A-X (0-0) | $R_2$(1/2) | 4300.31 | 39.1 |
| Na I | | | 5889.95 | < 669. |
| Na I | | | 5895.92 | < 607. |
| Li I | | | 6707.81 | 3.3 |
| K i | | | 7664.91 | 339. |
| K I | | | 7698.97 | 222. |

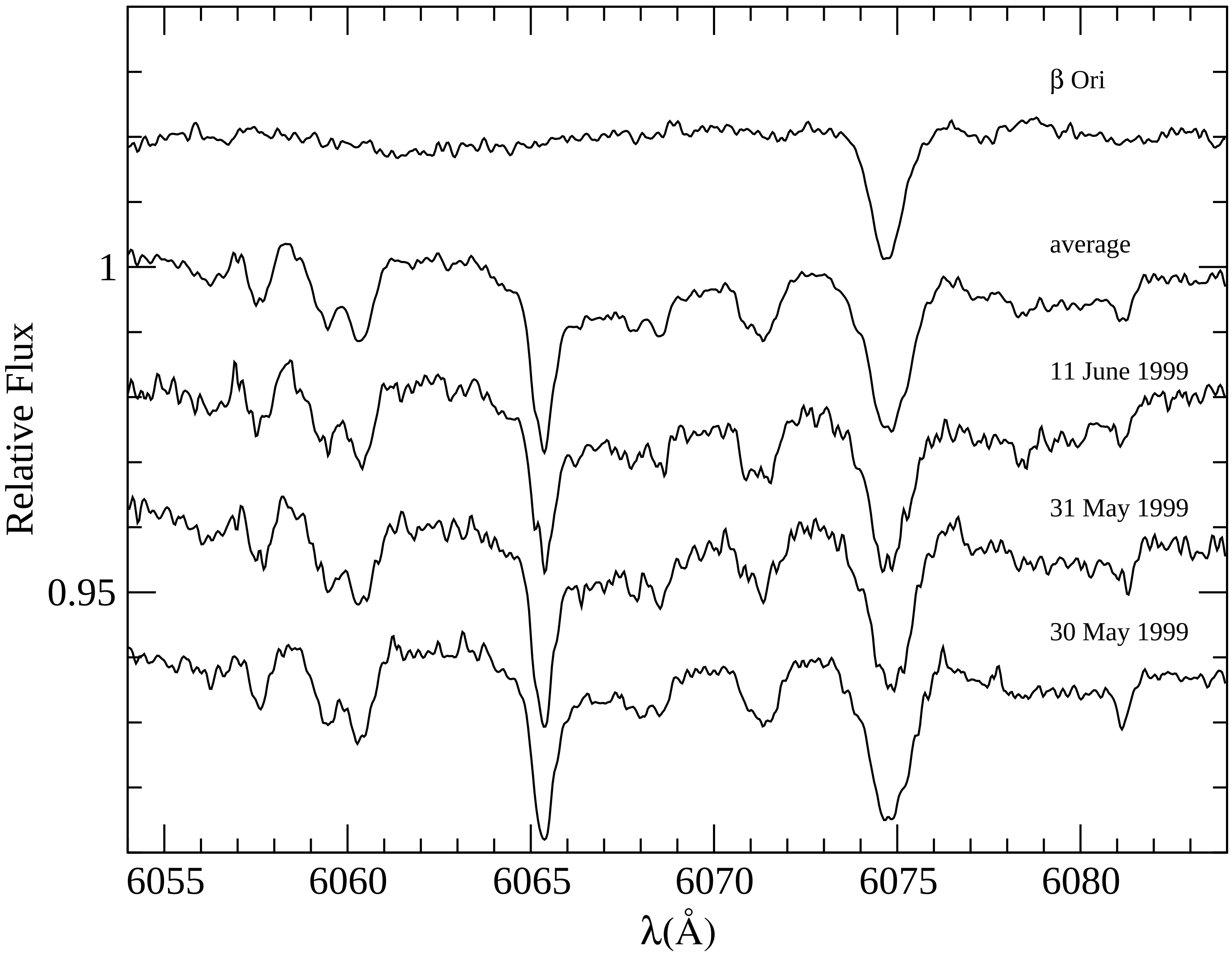

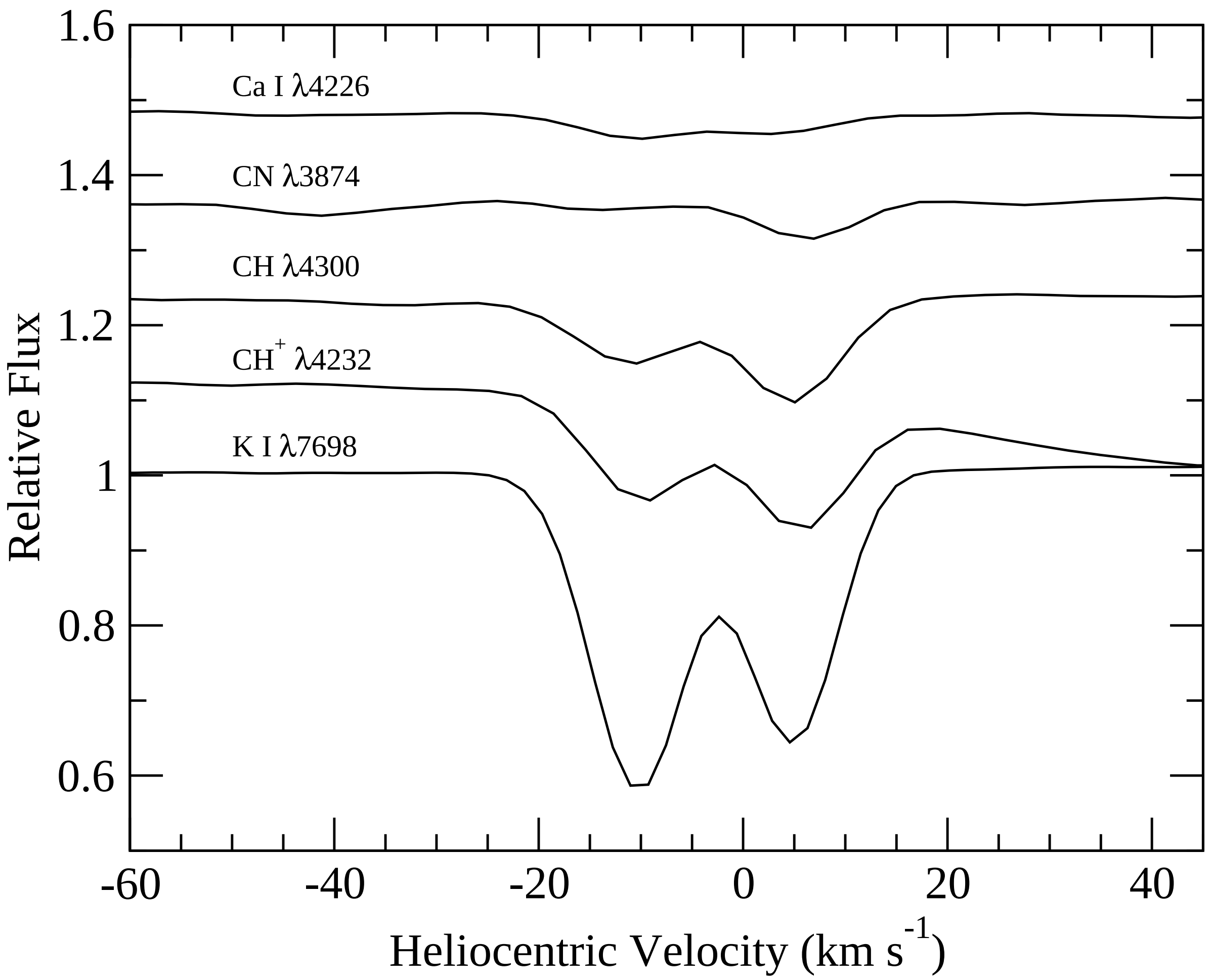

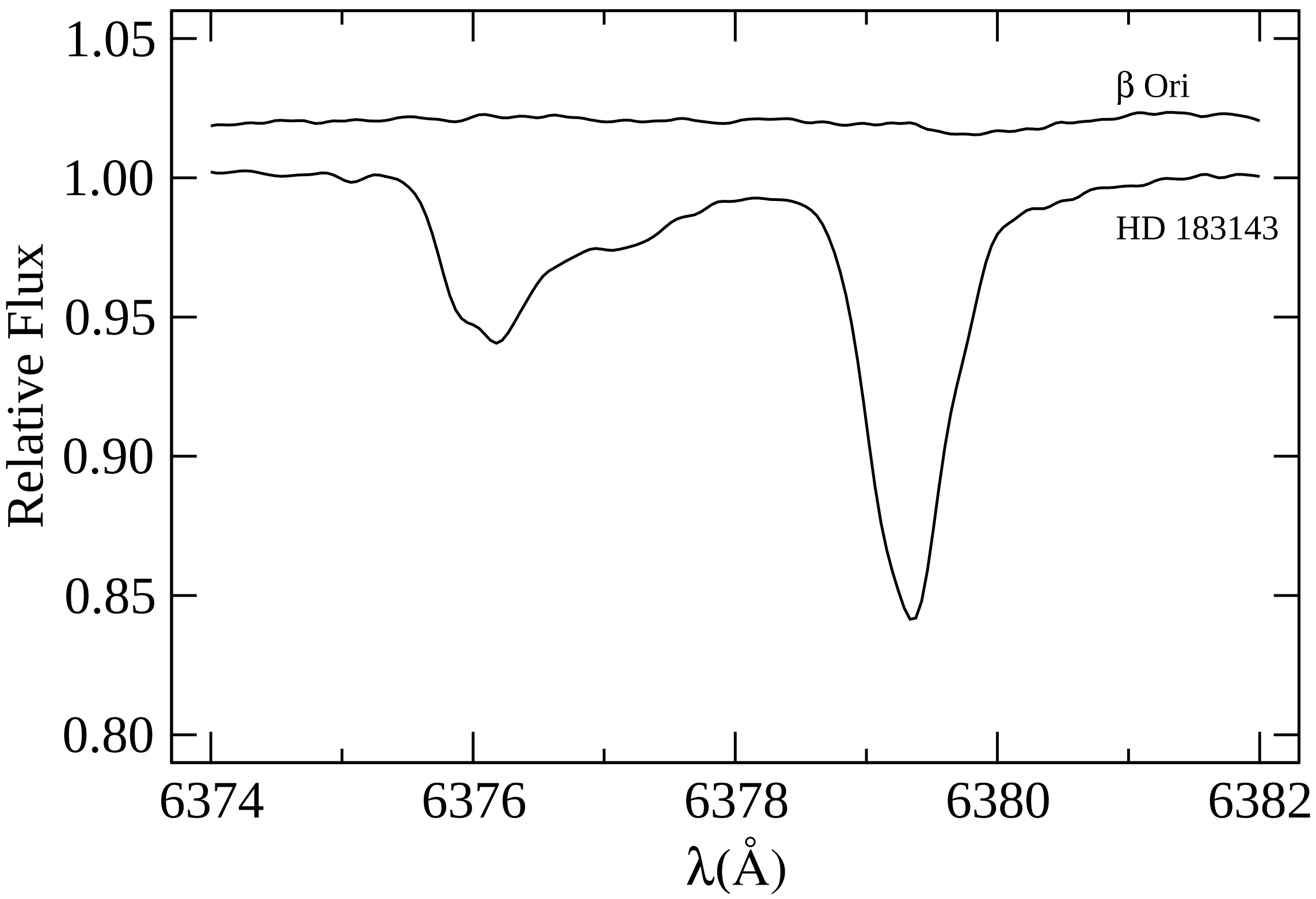

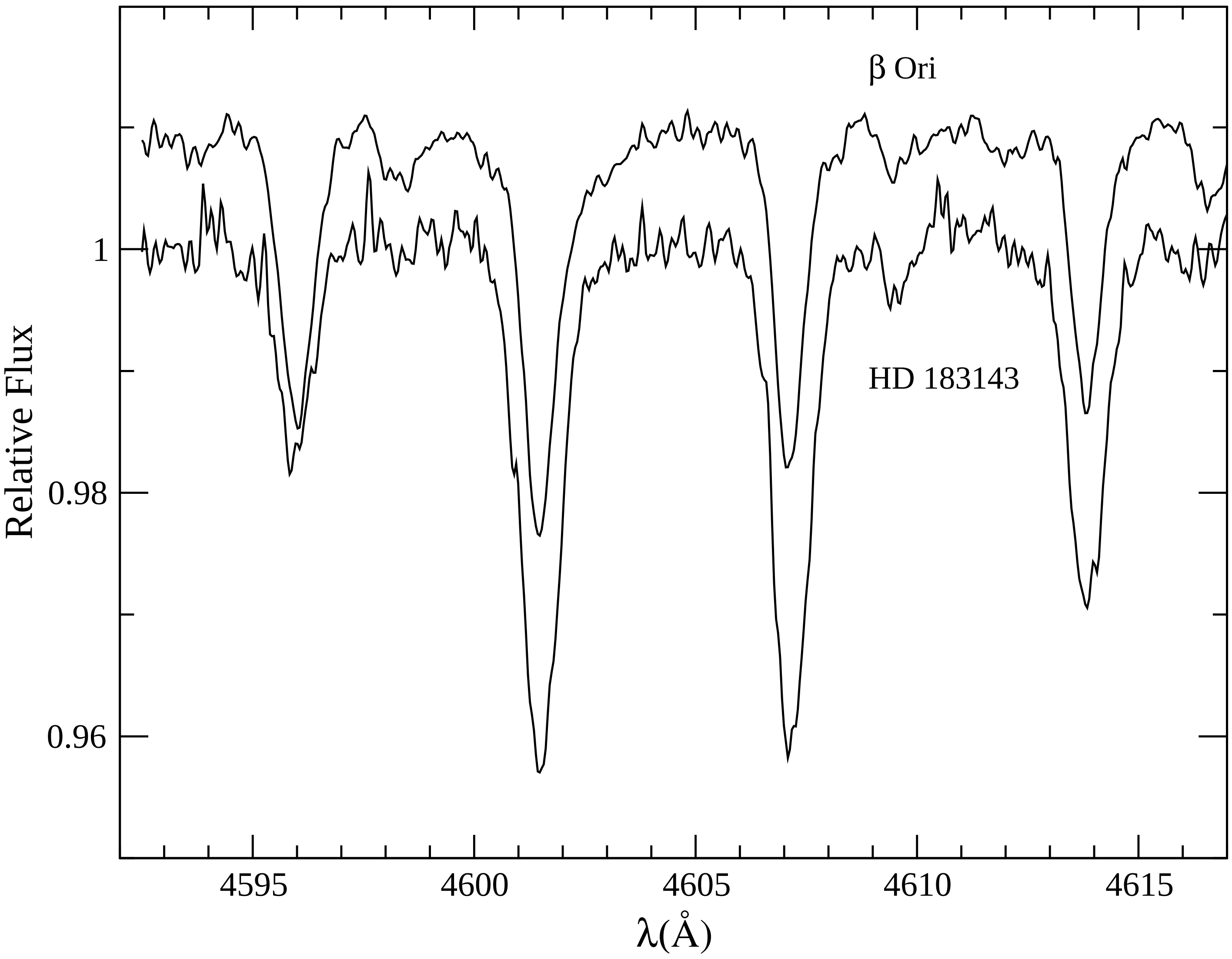

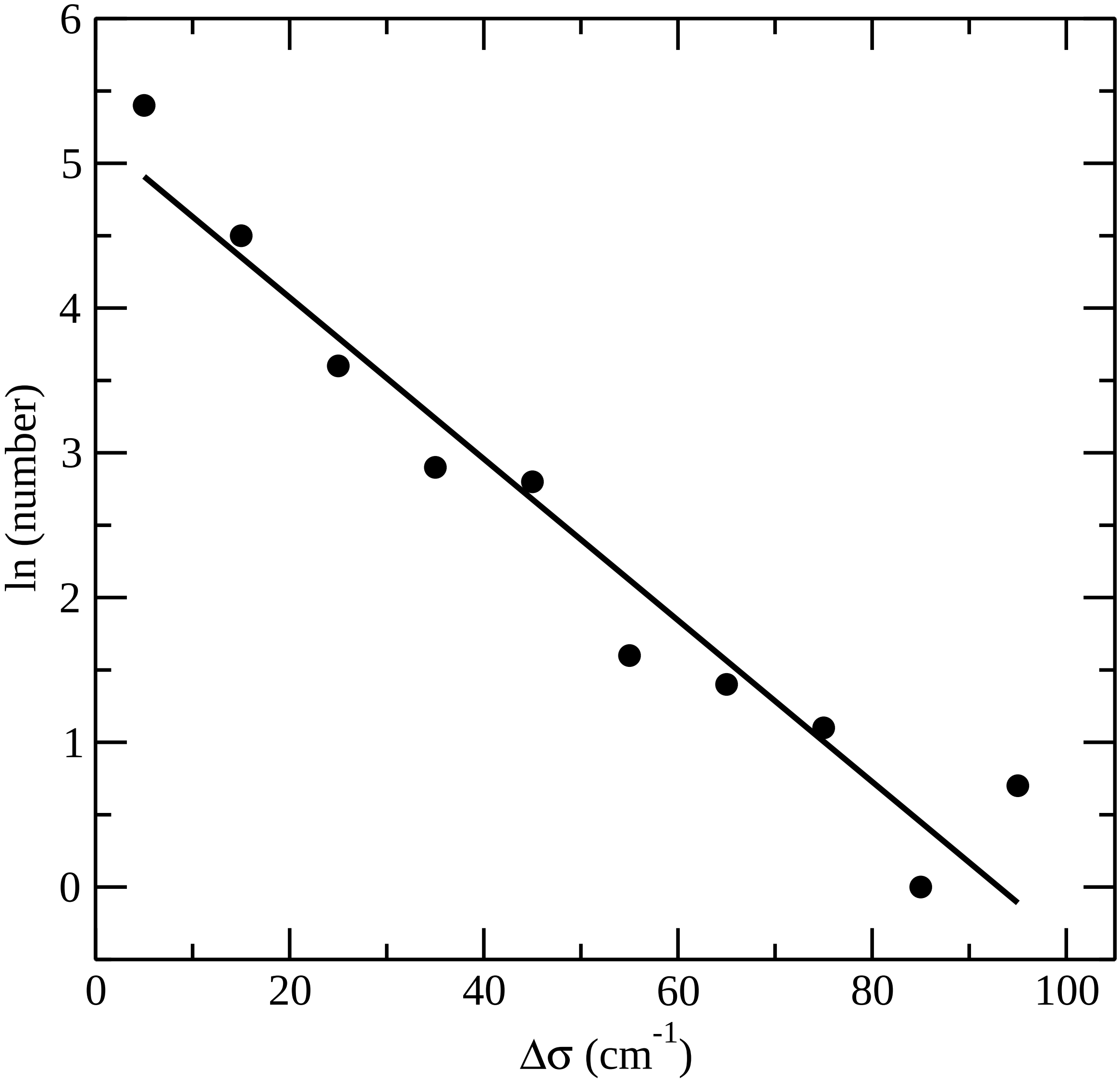

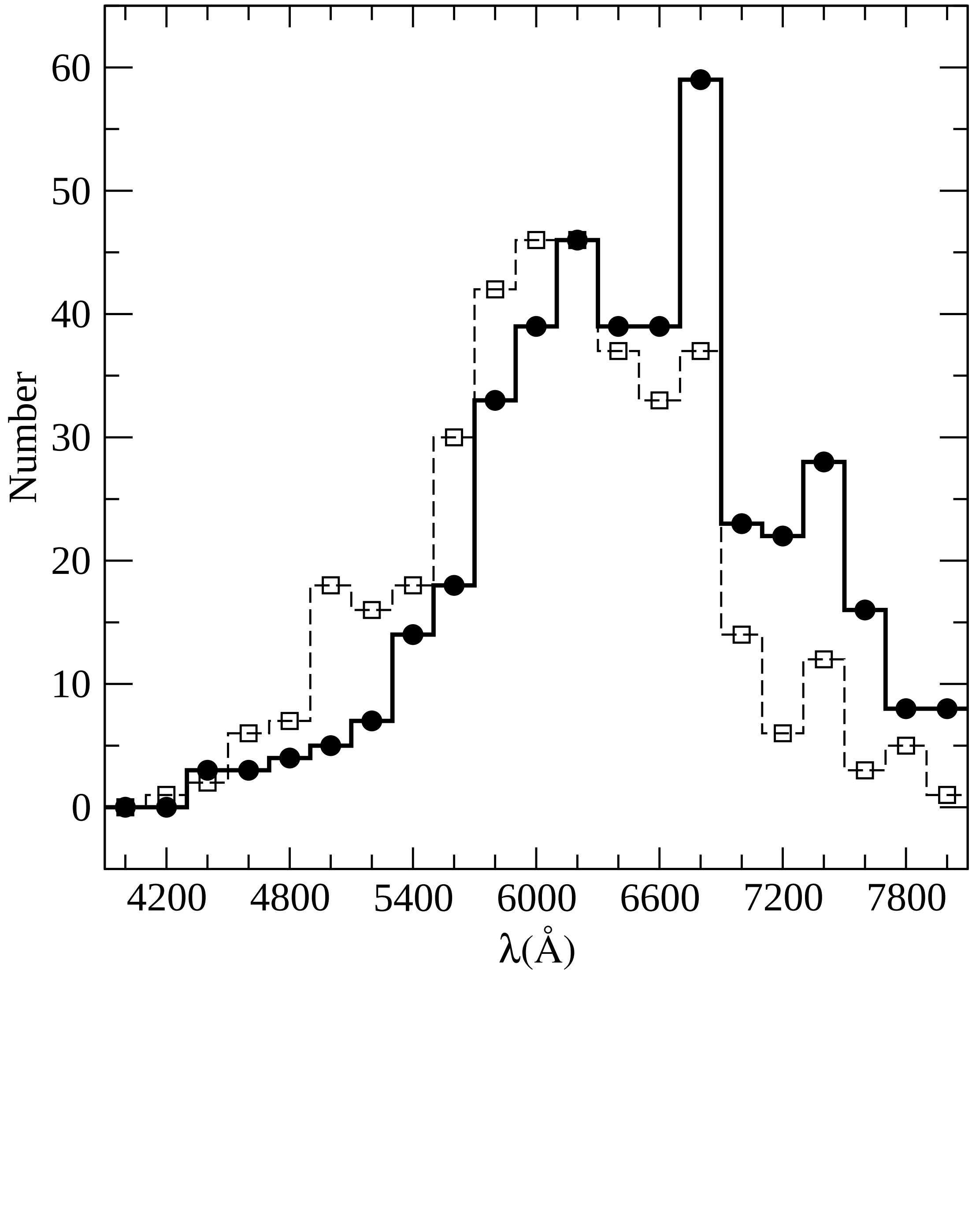

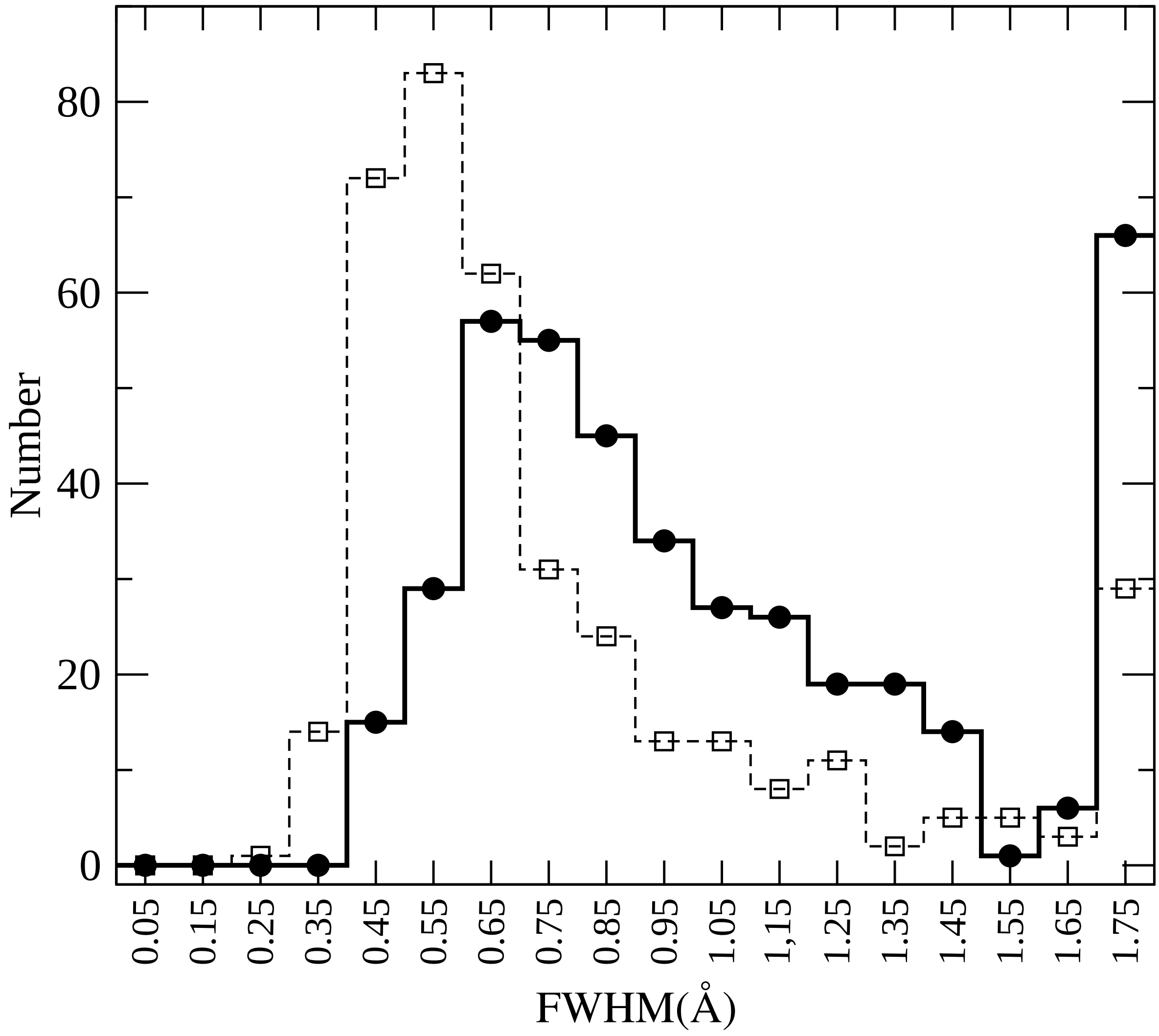

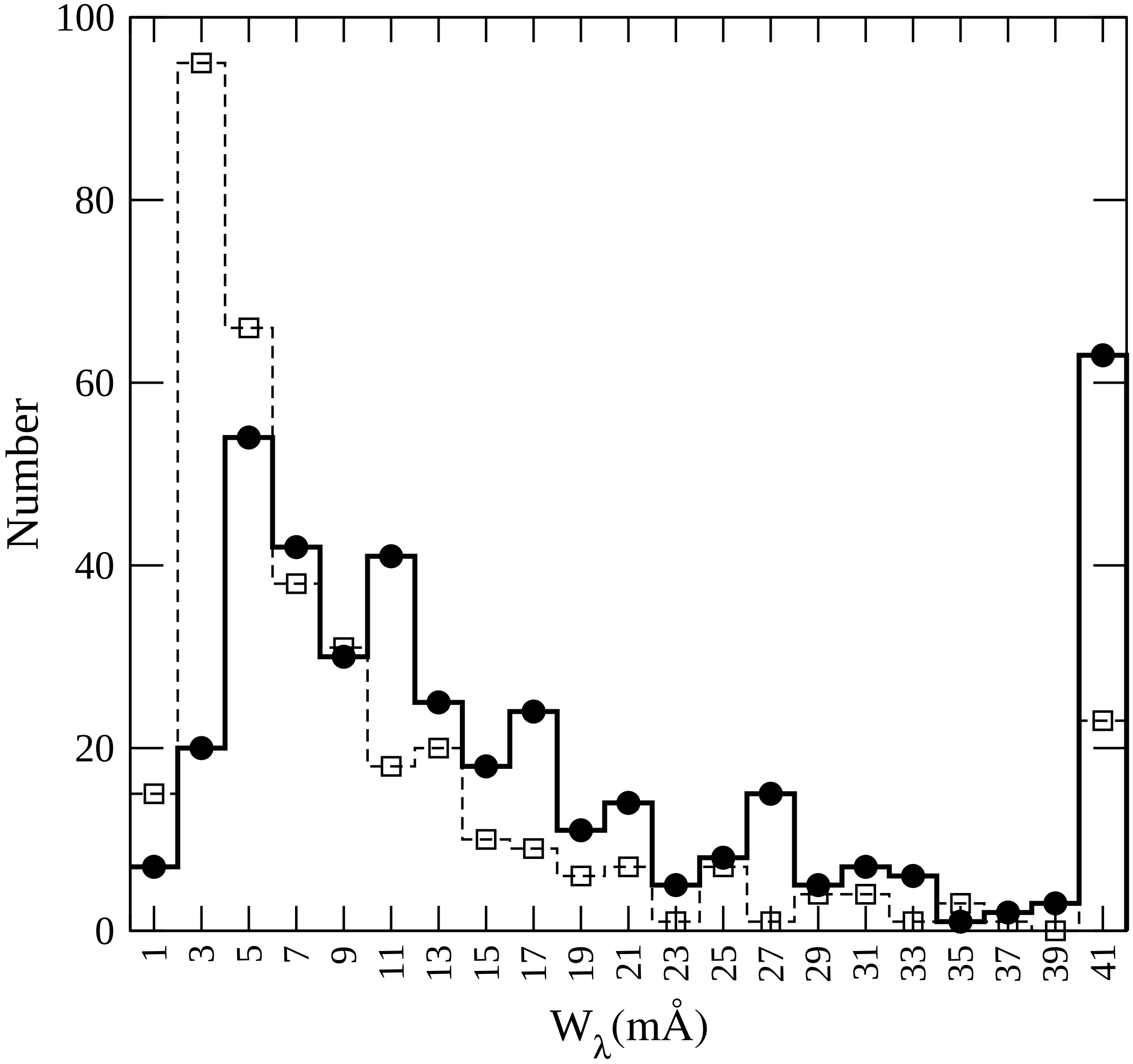

Number
0
20
40
60
80
100
1 3 5 7 9 11 13 15 17 19 21 23 25 27 29 31 33 35 37 39 41
$W_\lambda$(mÅ)

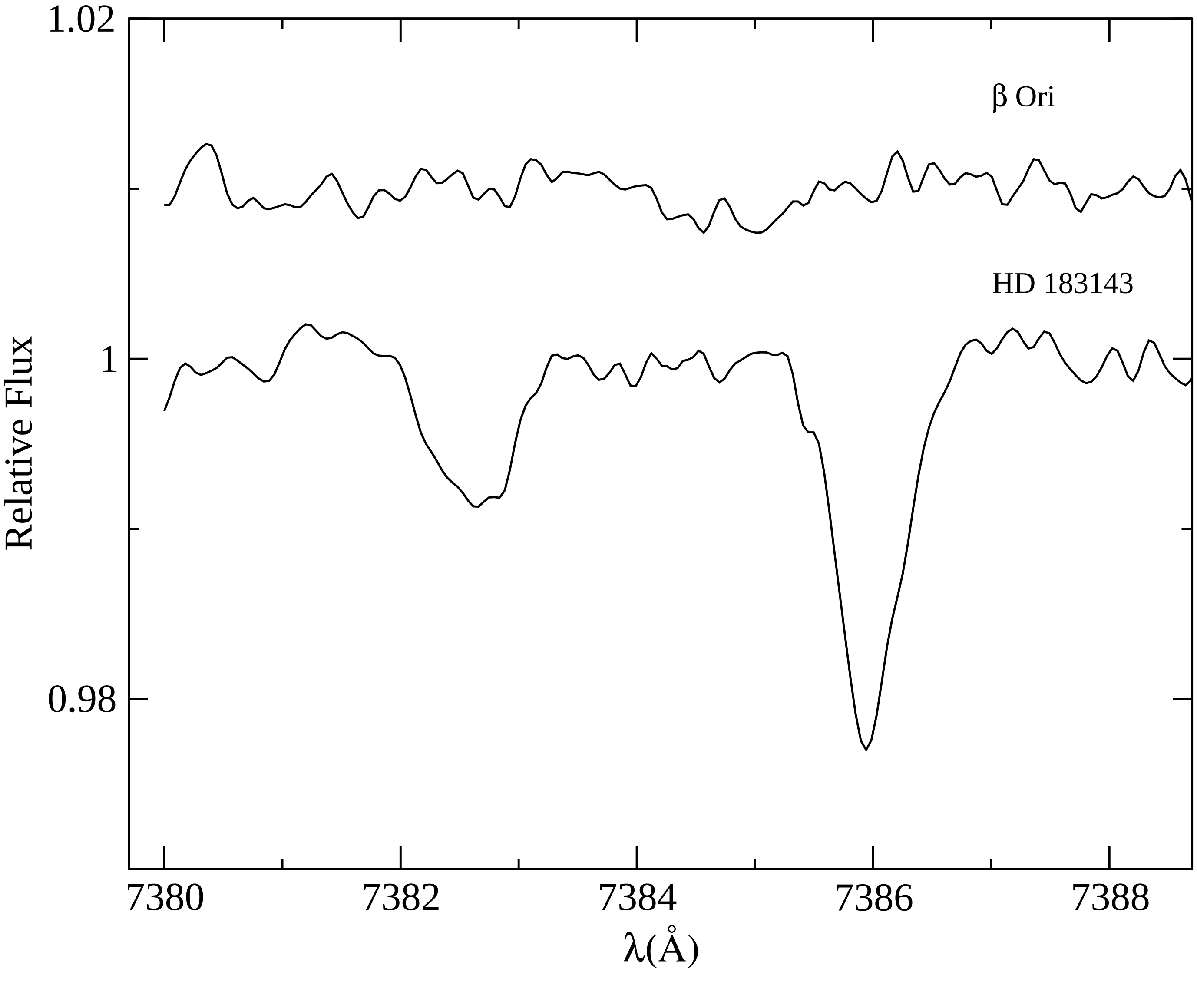

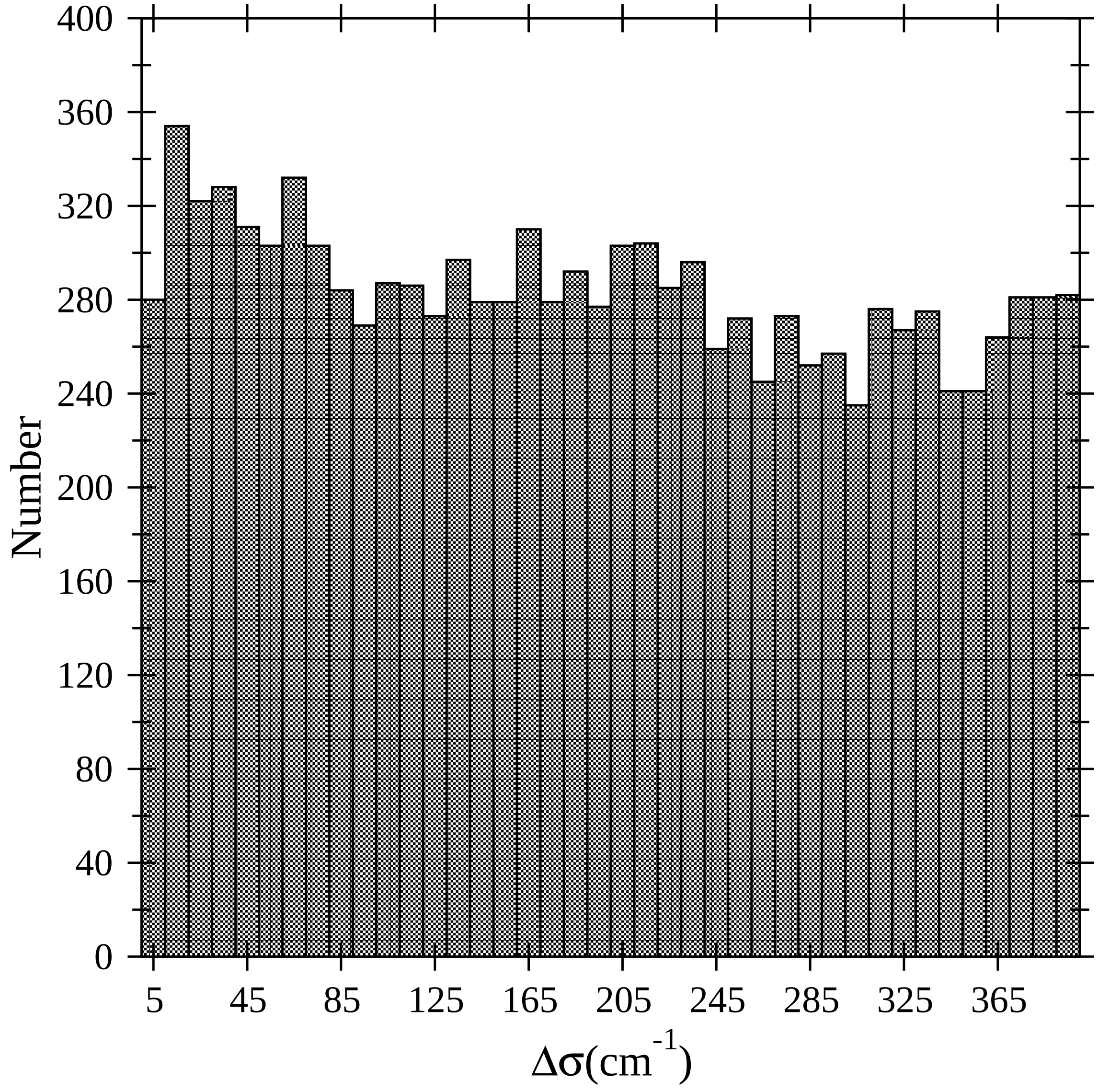

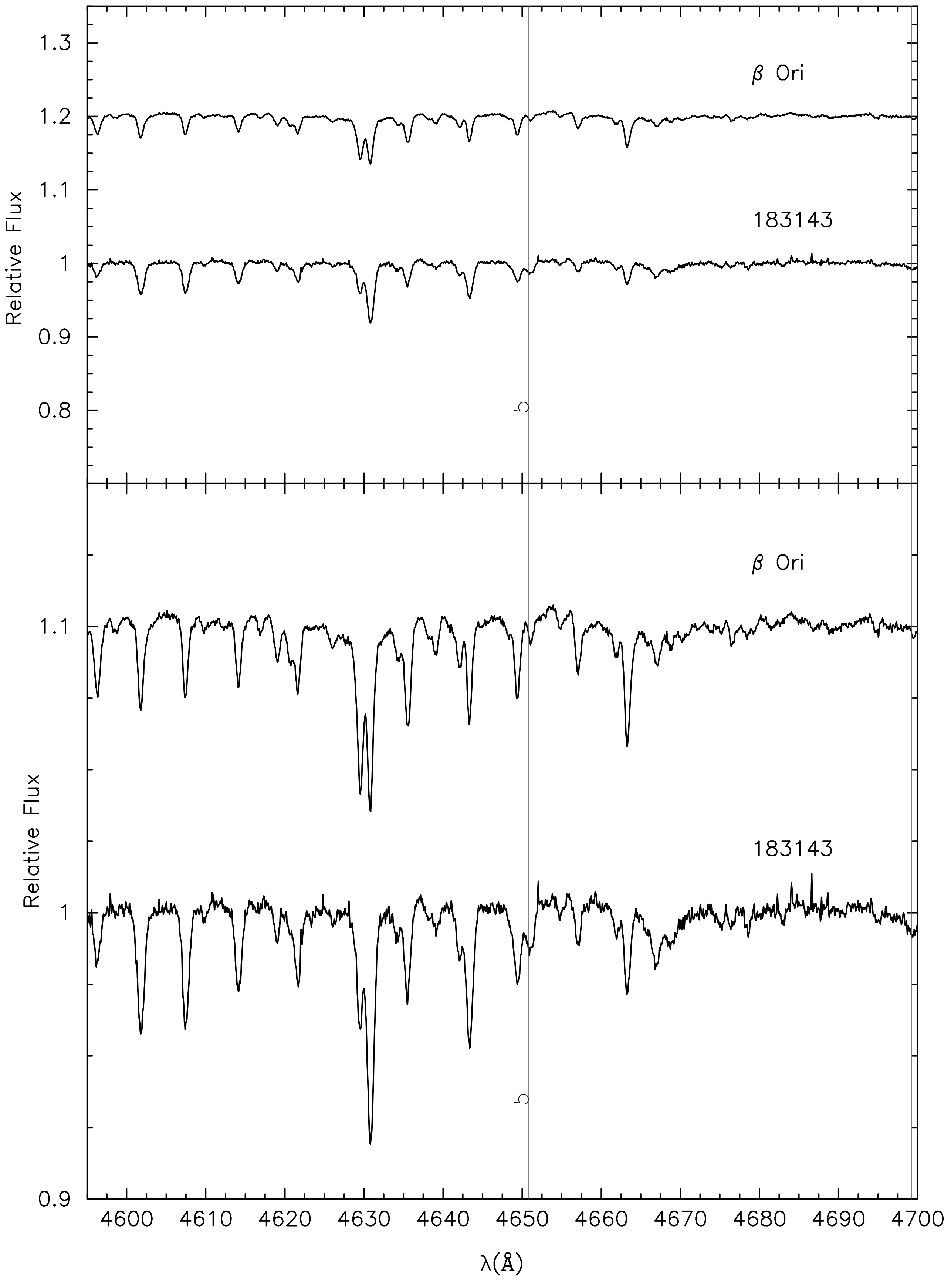

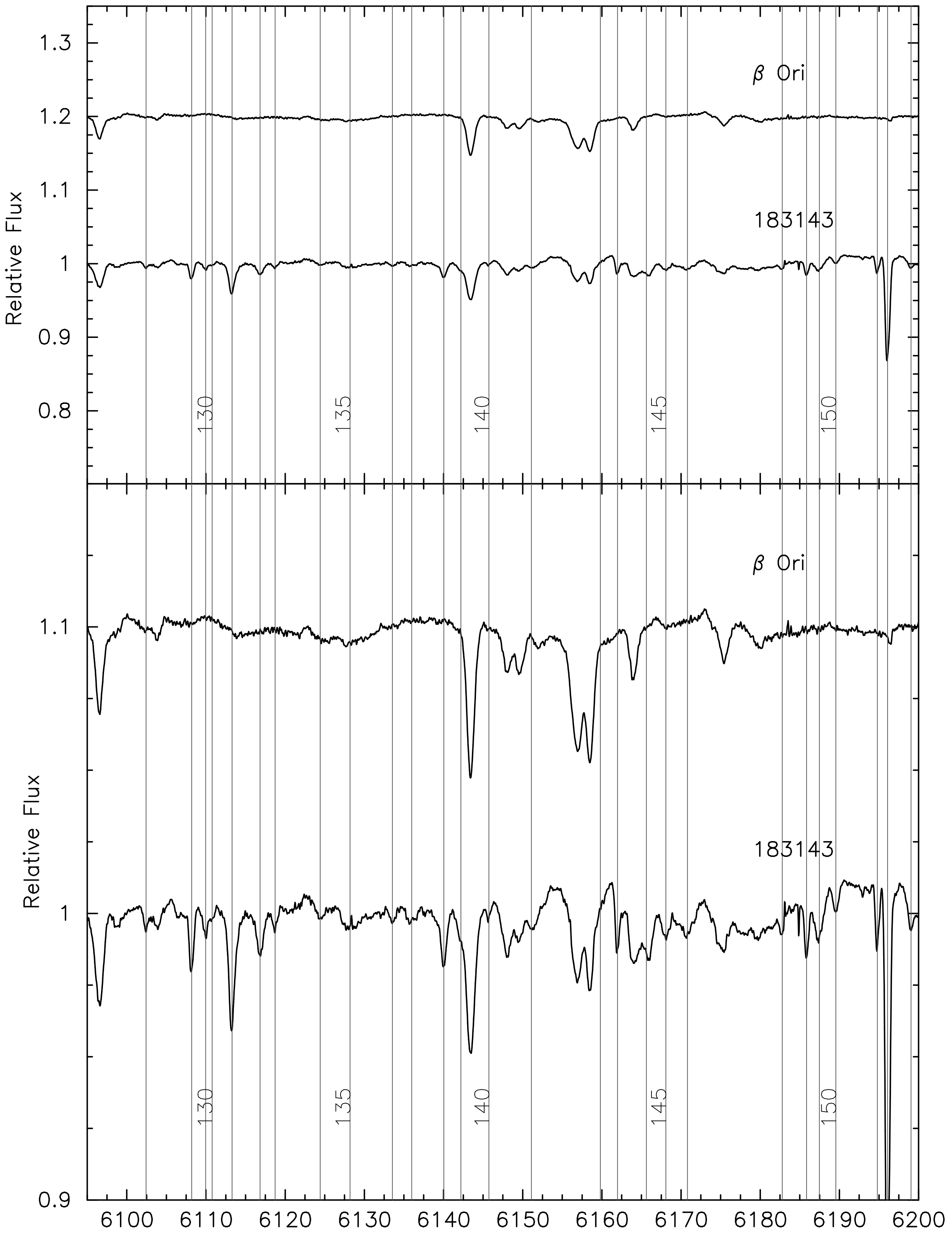

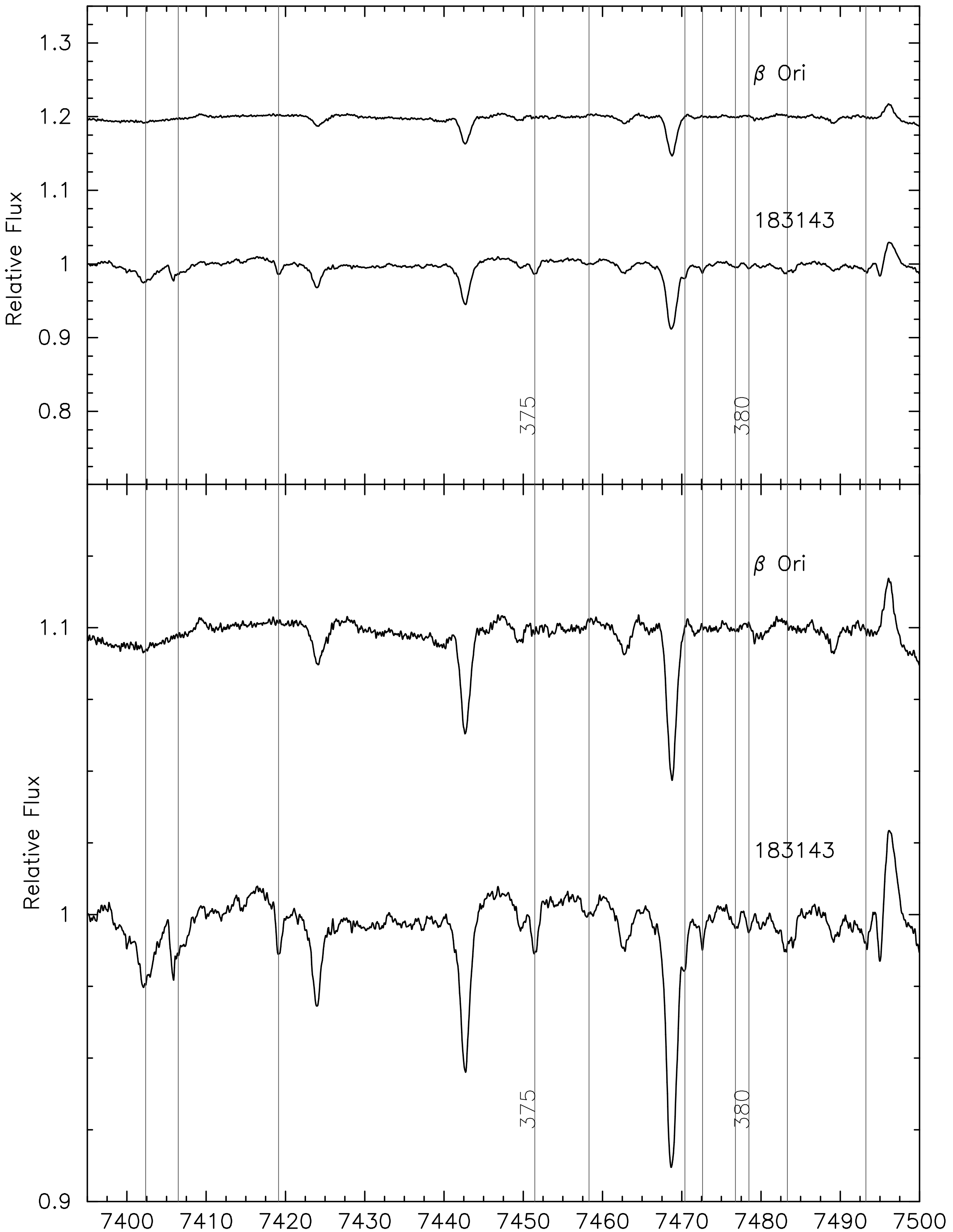